\documentclass{article}

\usepackage{PRIMEarxiv}

\usepackage[utf8]{inputenc} 
\usepackage[T1]{fontenc}    
\usepackage{hyperref}       
\usepackage{url}            
\usepackage{booktabs}       
\usepackage{amsfonts}       
\usepackage{nicefrac}       
\usepackage{microtype}      
\usepackage{lipsum}
\usepackage{fancyhdr}       
\usepackage{graphicx}       
\graphicspath{{media/}}     

\usepackage{multirow}
\usepackage{colortbl}
\usepackage{amsmath}
\usepackage{amssymb}
\usepackage{lipsum}
\usepackage{threeparttable}
\usepackage{balance}
\usepackage{enumitem}
\usepackage{longtable}
\usepackage[dvipsnames,svgnames]{xcolor}
\usepackage{adjustbox}
\usepackage{cite}
\usepackage{verbatim}
\usepackage{booktabs} 
\usepackage{multirow}
\usepackage[utf8]{inputenc}
\usepackage{url}
\usepackage{array} 
\usepackage{pifont}
\usepackage{wasysym}
\usepackage{utfsym}
\usepackage{fontawesome}
\usepackage{algorithmic}
\usepackage{graphicx}
\usepackage{stfloats}
\usepackage{textcomp}
\usepackage{lscape}
\usepackage{color}
\usepackage{hyperref}
\usepackage{subcaption}
\usepackage{soul, xcolor}
\usepackage{tabularx}
\usepackage{arydshln}

\title{MM-AttacKG: A Multimodal Approach to Attack Graph Construction with Large Language Models
}

\author{
  Yongheng Zhang \\
  College of Electronic Engineering \\
  National University of Defense Technology \\
  China \\
  \texttt{zhangyongheng@nudt.edu.cn} \\
   \And
  Xinyun Zhao \\
  College of Electronic Engineering \\
  National University of Defense Technology \\
  China \\
  \texttt{zhaoxinyun66@nudt.edu.cn} \\
   \And
  Yunshan Ma \\
  Singapore Management University Singapore \\
  Singapore \\
  \texttt{ysma@smu.edu.sg} \\
   \And
  Haokai Ma \\
  National University of Singapore \\
  Singapore \\
  \texttt{haokai.ma@u.nus.edu} \\
   \And
  Yingxiao Guan \\
  College of Electronic Engineering \\
  National University of Defense Technology \\
  China \\
  \texttt{guanyingxiao23@nudt.edu.cn} \\
   \And
  Guozheng Yang* \\
  College of Electronic Engineering \\
  National University of Defense Technology \\
  China \\
  \texttt{yangguozheng17@nudt.edu.cn} \\
   \And
  Yuliang Lu \\
  College of Electronic Engineering \\
  National University of Defense Technology \\
  China \\
  \texttt{luyuliang@nudt.edu.cn} \\
   \And
  Xiang Wang \\
  University of Science and Technology of China \\
  China \\
  \texttt{xiangwang@ustc.edu.cn} \\
}

\begin{document}
\maketitle

\begin{abstract}
Cyber Threat Intelligence~(CTI) parsing aims to extract key threat information from massive data, transform it into actionable intelligence, enhance threat detection and defense efficiency, including attack graph construction, intelligence fusion and indicator extraction.
Among these research topics, Attack Graph Construction~(AGC) is essential for visualizing and understanding the potential attack paths of threat events from CTI reports. 
Existing approaches primarily construct the attack graphs purely from the textual data to reveal the logical threat relationships between entities within the attack behavioral sequence.
However, they typically overlook the specific threat information inherent in visual modalities, which preserves the key threat details from inherently-multimodal CTI report.
Inspired by the remarkable multimodality understanding capabilities of Multimodal Large Language Models (MLLMs), we explore its potential in enhancing multimodal attack graph construction. 
To be specific, we propose a novel framework, MM-AttacKG, which can effectively extract key information from threat images and integrate it into attack graph construction, thereby enhancing the comprehensiveness and accuracy of attack graphs.
It first employs a threat image parsing module to extract critical threat information from images and generate textual descriptions using MLLMs. 
Subsequently, it builds an iterative question-answering pipeline tailored for image parsing to refine the understanding of threat images. 
Finally, it achieves content-level integration between attack graphs and image-based answers through MLLMs, completing threat information enhancement.
We construct a new multimodal dataset, AG-LLM-mm, and conduct extensive experiments to evaluate the effectiveness of MM-AttacKG. 
The results demonstrate that MM-AttacKG can accurately identify key information in threat images and significantly improve the quality of multimodal attack graph construction, effectively addressing the shortcomings of existing methods in utilizing image-based threat information. 
Code and the corresponding dataset will be released upon acceptance.
\end{abstract}

\keywords{Cyber Threat Intelligence \and Attack Graph Construction \and Multimodal Large Language Models}

\section{Introduction}
\label{introduction}


As cyber attacks increase in frequency and complexity, they represent a critical challenge to modern cybersecurity defenses.
Attack graph serve as the effective means to combat these escalating threats by graphically depicting the progression of attacks through interconnected nodes that represent individual attack steps, exploited vulnerabilities, and targeted assetss \cite{AttacKG,EXTRACTOR,ThreatKG,Attack,TTPDrill}.
Attack Graph Construction~(AGC), the task of generating such graphs systematically, relies on diverse data sources, such as system logs, human-curated knowledge, and Cyber Security Intelligence~(CTI) reports. 
Among these, CTI reports are especially promising. They deliver timely and precise threat intelligence, which helps identify key attack paths and focus on high-risk threats \cite{APT1,APT2,APT3,APT4}.
Due to its immense application value, the attack graph construction task has attracted extensive attention from both academia and industry.


\begin{figure}[h]
    \centering
    \includegraphics[width=0.8\linewidth]{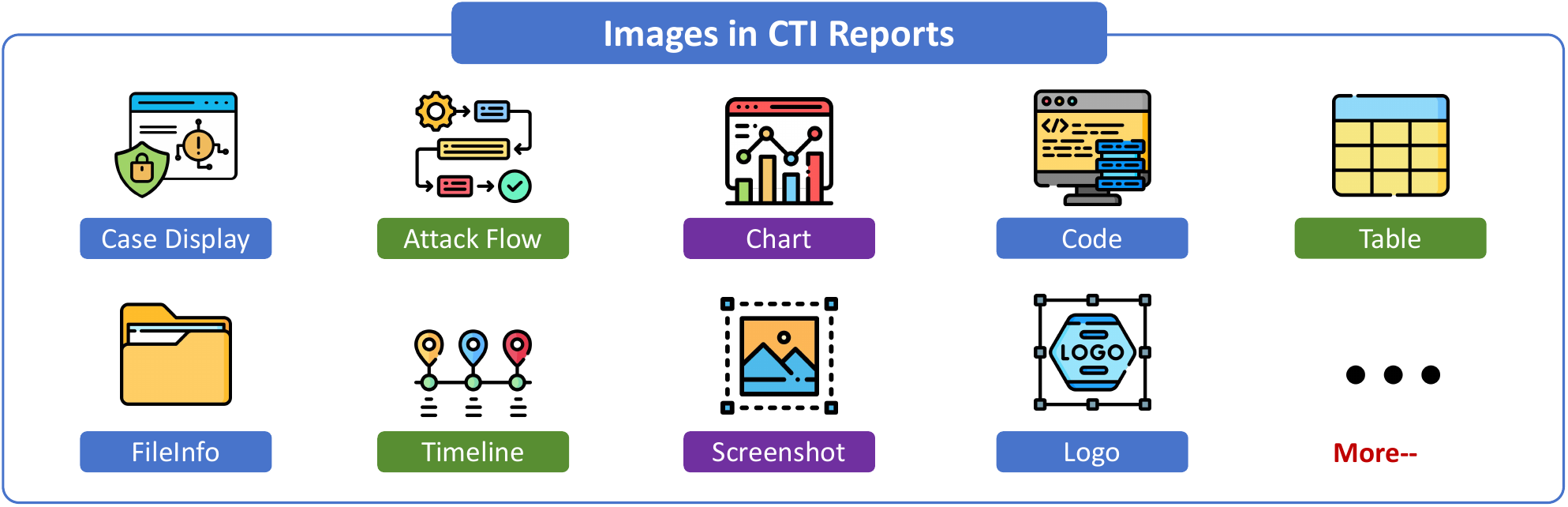}
    \caption{Motivation for incorporating images into constructing attack graphs is explained as follows. The images in CTI reports are highly complex and diverse, containing rich threat information. By parsing these images, we can greatly improve our understanding of threat events and thus enhance the quality of attack graphs.}
    \label{Motivation_case}
\end{figure}

The research on attack graph construction methods has gone through the following phases. 
In the initial phase, researchers established rules through expert knowledge, extracted Indicators of Compromise~(IoCs) based on regular expressions, and constructed attack graph frameworks as perceived~\cite{EXTRACTOR}.
However, this approach was limited by the fixed patterns of regular expressions and the subjective understanding of experts, making it difficult to adapt to the constantly changing tactics and techniques of cyber attackers. 
To address these challenges, deep learning-based methods have been proposed, such as AttacKG and ThreatKG~\cite{AttacKG,ThreatKG}.
Incorporating deep learning has significantly enhanced the construction efficiency of attack graphs. Nevertheless, the complexity of model selection and the stringent requirements on labeled data quality pose challenges for practical implementation. 
Then, with the tremendous success of Large Language Models~(LLMs), more and more researchers have begun to explore the use of LLMs to address the problem of attack graph construction. 
These pioneering works have explored ways to apply LLMs in attack graph construction tasks.
Such as the use of In-Context Learning (ICL) to apply LLMs to attack graph construction sessions like threat data automatic labeling~\cite{label}, threat entity recognition~\cite{entity}, and threat event extraction~\cite{AttacKG+}. 
Compared with traditional methods, LLM-based approaches demonstrate advantages in effectiveness, usability, and scalability. 

Despite the advantages of the LLM-based method, existing research often neglects the rich multimodal information in CTI reports, such as attack process diagrams, case study illustrations, and system screenshots. We refer to these visual materials containing threat-related content as "threat images".
As is illustrated in Figure~\ref{Motivation_case}, threat images exhibit complexity and diversity. Threat images are distributed in different locations of the CTI report according to their functional type, providing visual evidence of the corresponding attack scenarios, complementing the textual descriptions, and providing unique insights. Overlooking these images can limit the depth and accuracy of attack graph construction. How to parse threat images to enhance attack graphs remains a pressing research gap.



To bridge this gap, we aim to integrate images into cyber threat analysis to construct multimodal attack graphs.
Even though image understanding research is well established, we face three major challenges for this task~\cite{he2015deepresiduallearningimage,dosovitskiy2021imageworth16x16words,radford2021learningtransferablevisualmodels} : 
First, domain-specific knowledge is essential. Analyzing textual cybersecurity data requires domain knowledge to enhance LLMs' understanding of threat contexts~\cite{entity,AttacKG+}. Similarly, existing general LLMs lack inherent mechanisms to effectively parse cyber-specific visual semantics (e.g., network diagrams, intrusion detection alerts), limiting their applicability to threat image analysis~\textbf{(Challenge~1)}. 
Second, a picture is worth a thousand words. We need to determine how to precisely and swiftly capture the most critical information from threat images for attack graph construction. Therefore, a new prompt approach is needed to unleash the potential capabilities of MLLMs~\textbf{(Challenge~2)}.
Finally, the quality of threat information extraction is critical.
Previous studies~\cite{10.1145/3664647.3681593} have shown that improving performance through image feature exploration typically depends on large quantities of labeled training data. However, these data often lack generalizability across different task definitions. Consequently, there is a pressing need for a self-supervised mechanism capable of optimizing the extracted threat image information~\textbf{(Challenge~3)}.




To address the aforementioned challenges, we develop a novel framework for multimodal attack graph construction, named as~\textbf{MM-AttacKG}. 
Specifically, to solve the first challenge, we integrate cybersecurity knowledge into threat image parsing. In LLMs usage, prompt learning and knowledge introduction improve the expertise and relevance of threat image information parsing.
Then, to address the second challenge, we redefine threat image parsing as an iterative question-answering process, inspired by human brainstorming. Each question targets a specific aspect of attack graph construction.
During each iteration, questions are systematically generated to probe critical aspects of the image within the cybersecurity domain, ensuring targeted exploration of its meaningful implications. 
Finally, to address the third challenge, we set up two answer optimization paradigms. Evaluate the answer content from different dimensions and further optimize the parsed threat information to improve the quality of threat information extraction.
To evaluate our approach, we constructed an exploratory dataset by incorporating images from threat intelligence. 
We call the dataset \textbf{AG-LLM-mm}. 
The final evaluation results show that multimodal attack graphs have richer threat information than text-based attack graphs when supplemented with visual information. 
The main contributions are as follows:
\begin{itemize}[leftmargin=*, topsep=0.2pt,parsep=0pt]
      \item To the best of our knowledge, this is the first comprehensive study of exploring visual information for attack graph construction in the era of LLMs. 
      \item We designed a multimodal attack graph construction framework that identifies and integrates threat image information into an LLMs-based attack graph construction process.
      \item  Extensive experiments justify that our framework is able to effectively identify important information embedded in threat images, and the visual information can enhance the completeness of attack graph construction. In addition, these findings point out promising and relevant directions for future research.
\end{itemize}

\section{Related Works}
In this section, we review related work in three branches.1) CTI report extraction, 2) LLMs for cyber security. 3) Multimodal for cyber security.

\subsection{ CTI Report Extraction}
\textbf{Indicators of Compromise (IoCs)}.
Structured IoC sharing remains a cornerstone of open source CTI frameworks, as evidenced by platforms and studies such as~\cite{Threatcrowd,OpenCTI,AlienVault,Acing}. These systems catalog attributes including malicious file hashes, process identifiers, and malware metadata~\cite{Abuse, IBM,PhishTank}. However, their reliance on isolated, low-fidelity data limits their utility in reconstructing multi-phase adversarial campaigns, as they fail to capture contextual or behavioral linkages between indicators.

\textbf{Unstructured Text Intelligence Extraction}.
The cybersecurity community has developed advanced methods to transform unstructured threat reports into actionable intelligence.
Ramnan pioneered automated extraction of vulnerability exploitation patterns and mitigation strategies from unstructured CTIi~\cite{SemiAutomated}. Subsequent work by Ghazi established frameworks for correlating extracted threat concepts~\cite{Ghazi}, while Ghaith introduced information-theoretic metrics (entropy and mutual information) for text analysis in security contexts~\cite{Ghaith}.
The EXTRACTOR system~\cite{EXTRACTOR} generated attack behavior graphs without domain-specific text assumptions, complementing the joint extraction method of network relation triples from~\cite{CyberEntRel}.
Mao~\cite{inbook} addressed coreference resolution challenges in threat action extraction, whereas~\cite{ATDG} mitigated OOV issues through multi-granular feature extraction. Advanced neural architectures like TA-GCN~\cite{SHANG2024111829} and KnowCTI~\cite{WANG2024103824} further improved entity-relationship modeling by integrating domain knowledge and dependency-aware embeddings.

\textbf{TTP Identification \& Operationalization}.
MITRE Technology, Tactics, Procedures~(TTP) matrix~\cite{Mitre} has become the de facto classification standard for cyber threat adversarial pattern recognition. Recent advances include:
Context-aware TTP extraction via~\cite{TTPDrill} enables real-time defense orchestration, while Ayoade automates TTP classification and mitigation mapping from heterogeneous threat reports~\cite{Ayoade}.
Ge employs semantic impact scoring with conditional probability for TTP prediction~\cite{SeqMask}, whereas Liu's ATHRNN architecture captures hierarchical TTP dependencies through transformer-recurrent hybrid networks~\cite{Liu}.
The TCENet framework bridges TTP and operational defense by generating Sigma rules directly from TTP descriptions~\cite{You}.

\subsection{Multimodal for cyber security}
Multimodal learning enhances threat understanding by fusing heterogeneous data sources.

\textbf{Multimodal data fusion and threat detection}.
In the area of cyber threat analysis and modeling, Nirnimesh proposed a multimodal graph-based approach for modeling and analysing cyber attacks~\cite{Ghose}. This work constructs a comprehensive analysis framework that includes victims, adversaries, autonomous systems, and cyber events by representing the stages, actors, and outcomes of cyber attacks as multimodal graphs. 
Multimodal data fusion techniques were widely used for critical infrastructure protection and threat detection. For example, Nikolaos proposed an attack detection framework based on multimodal data fusion and adaptive deep learning for the vulnerability of critical water infrastructure to cyber attacks~\cite{8653521}. The framework improves the detection of attacks by integrating multimodal data. In addition, Li proposed an LLM-based detection method for phishing attacks, which combines image and text information of web pages to overcome the limitations of traditional single-modal-based methods~\cite{299740}.

\textbf{Multimodal threat information recognition}.
Automated extraction of cyber threat intelligence is another important application area of multimodal learning. Zhang proposed the EX-Action framework, which extracts threat actions from CTI reports through natural language processing techniques and combines them with multimodal learning algorithms to identify threat behaviors~\cite{Huixia}. Similarly, Xiao proposed an advanced persistent threat participant attribution method based on multimodal and multilevel feature fusion (APT-MMF), which solves the problem of ignoring heterogeneous information in existing methods~\cite{Xiao}.

\textbf{Security of Multimodal Large Language Models}.
To address the security of multimodal learning models, researchers have proposed various defense mechanisms. Liu proposed a ‘machine forgetting’ based defence mechanism for backdoor attacks in multimodal comparative learning to reduce the impact of malicious behaviors on the inference process of the model~\cite{Kuanrong}. Shan reveals the vulnerability of text-to-image model generation to poisoning attacks under large-scale training data and proposes a ‘night shadow’ attack method to precisely control the model output through a small number of poisoned samples~\cite{Shawn}. JailGuard focuses on the vulnerability of large and multimodal language models to jailbreak and hijacking attacks and proposes a generic detection framework for identifying multiple attacks across modalities~\cite{zhang2024jailguarduniversaldetectionframework}.

\subsection{LLMs for Cybersecurity Threat}
The application of large language models in cybersecurity has emerged as a multi-faceted research frontier.

Systematic methodological frameworks form the theoretical basis for LLM security implementations. Kucharavy established essential taxonomies for generative language modeling, delineating capability boundaries and implementation principles~\cite{kucharavy2023fundamentals}. Building on this foundation, Wrsch proposed a semantic pattern recognition framework for cybersecurity knowledge entity extraction~\cite{wrsch2023llms}, while Pan achieved breakthroughs in log anomaly detection through retrieval-augmented architectures with vector database integration~\cite{pan2023raglog}.
In complex pattern recognition, Ferrag demonstrated FalconLLM's effectiveness in identifying multi-stage attack vectors through contextual reasoning~\cite{ferrag2023revolutionizing}. Charan quantitatively evaluated LLMs capabilities in implementing MITRE ATT\&CK techniques, establishing benchmark comparisons between ChatGPT and Bard for tactical code generation~\cite{charan2023text}.
For security operations, Rigaki deployed pre-trained models as autonomous agents in network environments, optimizing sequential decision-making processes~\cite{rigaki2023cage}. The chained prompt engineering framework proposed by Moskal structured threat response workflows through plan-act-report cycles~\cite{moskal2023llms}.
Addressing LLM domain adaptation challenges, Kereopa constructs an evaluation system combining expert case analysis with quantitative metrics, while current research generally faces the challenge of enhancing models' cybersecurity-specific cognition~\cite{kereopayorke2023building}.

\section{Preliminary}
\label{Preliminary}

\subsection{Problem Formulation}
\label{Problem Formulation}
Current research on multimodal attack graph construction does not have a rigorous paradigm.
Specific application scenarios influence the characterization and structure of attack graphs. 
To formally define the problem, we adopt the attack graph definition in text-based attack graph work as a basis.


\textbf{CTI Report}. A CTI Report is an evidence-based, structured analysis that encompasses the context, mechanisms, indicators of compromise, potential impacts, and actionable recommendations regarding existing or emerging cyber threats. It serves as a critical output of the cyber threat intelligence lifecycle, providing organizations with actionable insights to proactively manage and mitigate cyber risks.

\textbf{Attack Graph}. An attack graph is a graphical representation method used for modeling and analyzing cyber attack pathways. It illustrates the relationships of the elements in a threat event and the steps taken by an attacker to launch an attack, thereby assisting security analysts in understanding potential threats within a network environment and assessing the security posture of a system.

\begin{figure}[!t]
    \centering
    \includegraphics[width=0.95\linewidth]{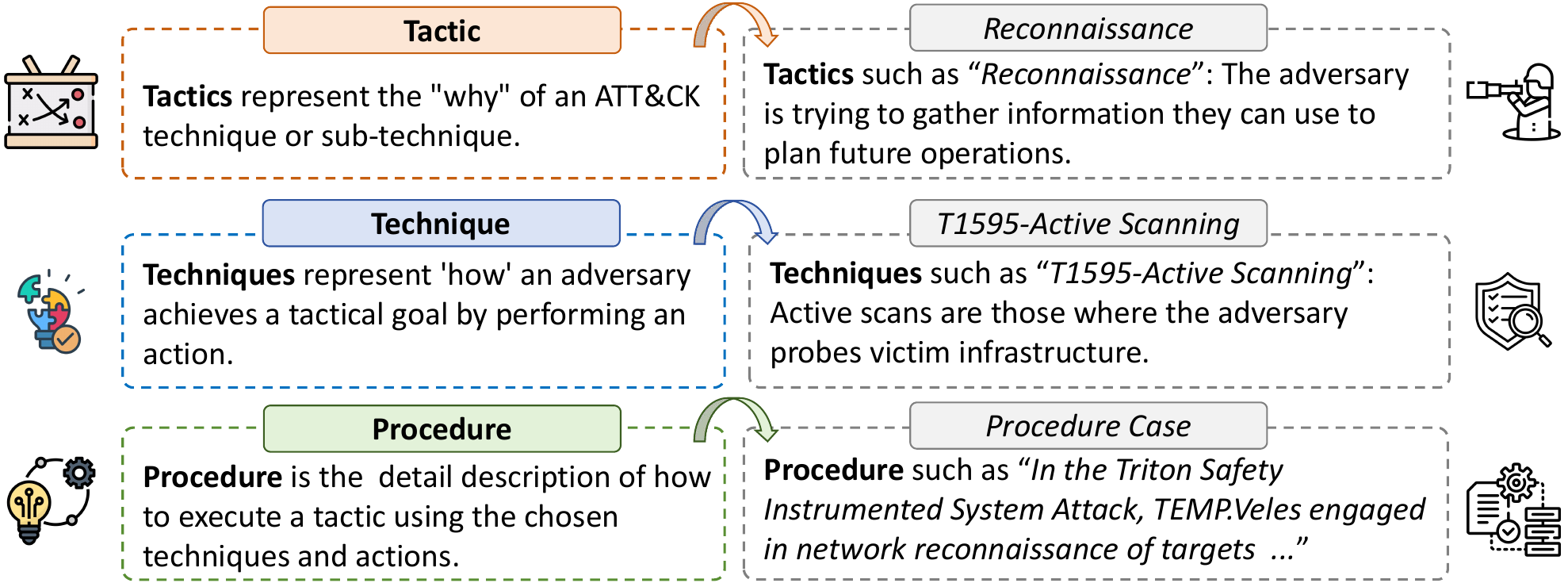}
    \caption{The atomic event data structure of the attack graph contains three parts: \textbf{Tactics}, \textbf{Techniques} \& \textbf{Procedures}, where procedures are presented in the structure of the threat behavior graph.}
    \label{attack_graph_tree}
\end{figure}

\textbf{Text-based Attack Graph Construction}. Text-based attack graph obtains threat information from the textual modal content of CTI reports. Given a collection of historical threat intelligence text events $E=\left\{e_1, e_2, \ldots, e_k\right\}$. This formulation defines extracting each threat behavior as a quadruple~$(s, a, o, t)$, which is also called an atomic event, where~$s, a, o$ and $t$ correspond to the subject, threat action, object, and timestamp in the current threat event. 
At each timestamp $t$, all the quadruples form an event graph, denoted as~$AG_{t}=\{(s, a, o, t)\}^N$, where $N$ is the number of atomic events at the threat event.  
It can be seen that the scheme uses a sequence of threat actions to characterize the evolutionary process of the attack graph.
Furthermore, as is shown in Figure~\ref{attack_graph_tree}, some works~\cite{AttacKG} focus on the content support of TTP~(Technology, Tactics, Procedures) labels for attack graphs to characterize the cyber threat technologies to which threat atomic events are mapped.
Specially, each atomic event is extended from a quadruple~$(s, a, o, t)$ to a quintuple $(s, a, o, t, p)$, where $s \in \mathcal{E}, a \in A, o \in \mathcal{E}$ and $p \in P$, represent the subject, threat action, object, and TTP label.
Correspondingly, the atomic event at each timestamp will be extended as~$AG_{t}=\{(s, a, o, t, p)\}^N$. 
It is worth mentioning that for the presence of non-verbal relations $R$ in threat events, $R$ is the set of non-verbal relations that are not ignored, but rather are used as attack graph supplementary links that hang over the relevant entities, but not as attack steps descriptions.

\textbf{Multimodal Attack Graph Construction by Image-enhanced}.
Based on the text-based attack graph, the images associated with structured events in CTI reports are introduced to construct the multimodal attack graph, where the images are represented by $V=\left\{v_1,v_2,\ldots,v_m\right\}_{m=1}^M$, where $M$ is the number of images. The image-enhanced attack graph construction process is divided into two phases: multimodal attack information extraction and attack graph integration.
\begin{itemize}[leftmargin=*, topsep=0.2pt,parsep=0pt]
      \item \textbf{Image Attack Information Extraction}. 
      This phase is based on the content summarization of the text-based attack graph, and parses the attack information contained in the images of the CTI report from different aspects, providing additional context and detail for the construction of the attack graph.
      \item \textbf{Multimodal Attack Graph Integration}. 
      This phase leverages the attack information extracted from the images in the CTI report as the basis, and supplements the text-based attack graph from three dimensions: entities, relation, and techniques. By integrating threat information from multiple modalities, it forms an image-enhanced attack graph.
\end{itemize}

Finally, the multimodal attack graph construction task can be characterized as follows: given a collection of historical threat intelligence text events $E=\left\{e_1, e_2, \ldots, e_k\right\}$ and associate images~$V=\left\{v_0, v_1, \ldots, v_{m}\right\}$, identify quintuples~$(s, a, o, t, p)$ from text and images form that can portray the evolutionary flow of threat events to form a multimodal attack graph.

\subsection{Text-based Attack Graph Construction}
\label{Textual Attack Graph Constructor}
AttacKG+~\footnote[1]{\url{https://github.com/multilayer-go/AttacKG-plus}} is a novel attack graph construction framework based on large language models, designed to transform textual cyber threat intelligence reports into structured attack graphs. 
The framework employs a modular design to build multilayered attack graphs, comprising four modules: Rewriter, Parser, Identifier, and Summarizer. Each module leverages the instruction prompting and in-context learning capabilities of LLMs to sequentially perform tasks such as report rewriting, behavior graph extraction, technique label matching, and state summarization. 
Furthermore, AttacKG+ introduces an upgraded attack knowledge schema that represents the attack process as a temporally unfolding complex event. 
Each temporal step encompasses three layers: behavior graph, TTP labels, and state summary, providing a more comprehensive characterization of the attack process.

\begin{figure*}[!t]
    \centering
    \includegraphics[width=0.98\linewidth]{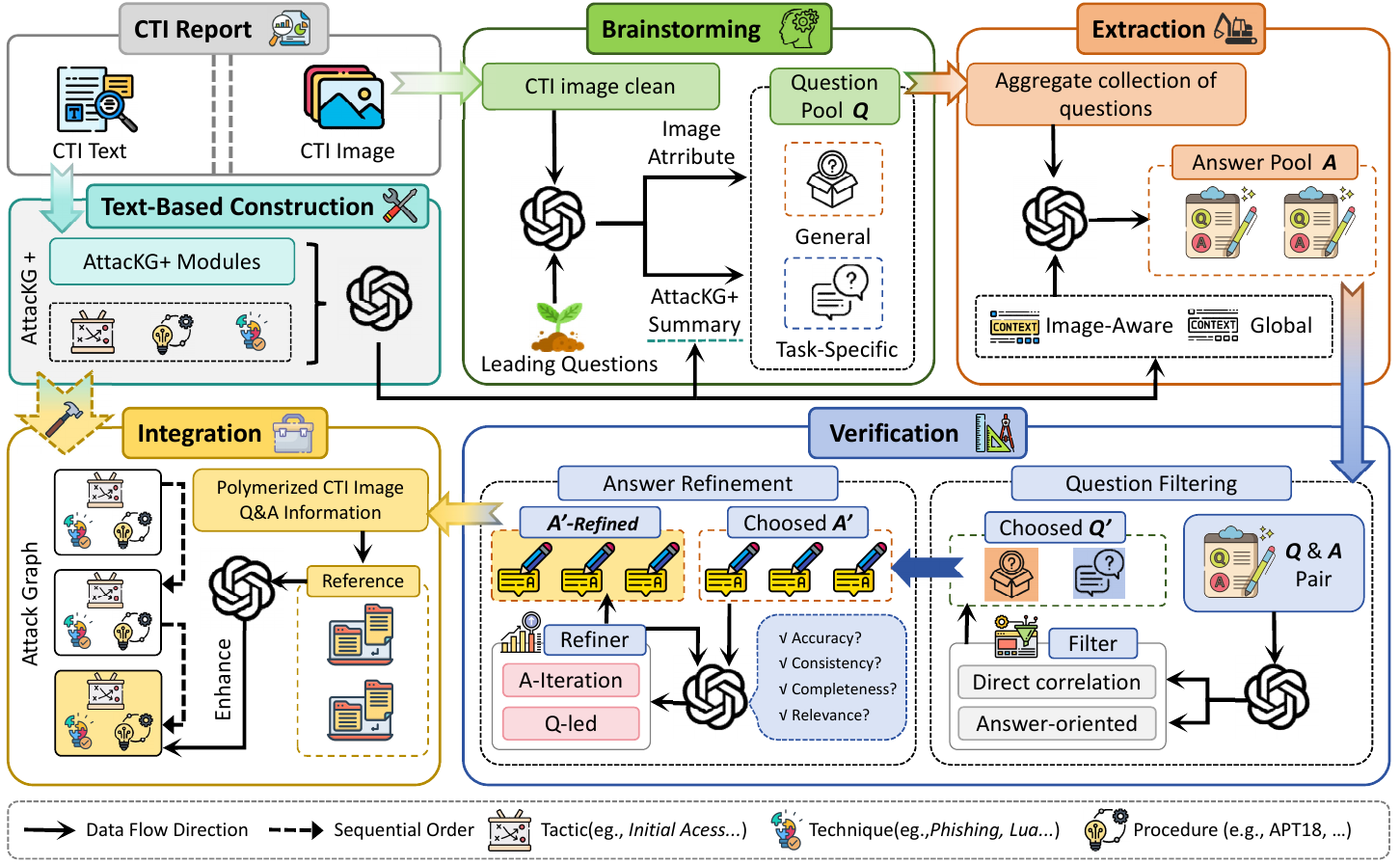}
    \caption{The overall framework of MM-AttacKG consists of five modules: (1)~\textbf{Text-based Construction}, here we use AttacKG+ as the constructor to construct a text-based attack graph by parsing CTI text. (2)~\textbf{Brainstorming}, the module targets key aspects of image parsing through question generation; (3)~\textbf{Extraction}, the module that combines the question set with the image content to extract key information from CTI images; (4)~\textbf{Verification}, the module optimizes the quality of threat image understanding through a two-stage process of question filtering and answer refinement; and (5)~\textbf{Integration}, the module refines the attack graph by adding modal information from CTI image.}
    \label{overview}
\end{figure*}

By utilizing the powerful language understanding and zero-shot learning capabilities of LLMs, AttacKG+ overcomes the limitations of traditional methods in terms of generalization ability and adaptability to new attack scenarios. 
Given the superior processing performance of AttacKG+ in text-modal threat intelligence, we use it as a textual attack graph constructor for multimodal attack graph construction work as shown in Figure~\ref{overview}.
Meanwhile, we obtain the summary information of the current report~(Global-Context) and the image context information~(Image-Aware-Context) as the parsing support of the threat image through the summary function module in it.

\section{Approach}
\label{Approach}

The proposed framework is outlined in Figure~\ref{overview}.
Our approach MM-AttacKG consists of five phases, which smoothly implements the flow from threat image parsing to attack graph multimodal gain. 
Section 4.1 outlines the brainstorming procedure, which is intended to clarify the key aspects of threat image parsing.
Section 4.2 defines the threat information extraction process, which is specifically designed to accurately extract threat-related information from threat images.
Section 4.3 presents the question filtering mechanism, which can effectively screen out threat information aspects that meet the criteria for constructing attack graphs.
Section 4.4 proposes two optimization paradigms to enhance the quality of answers in threat image extraction.
Section 4.5 introduces the attack graph integration scheme, which can integrate the parsed threat image information into text-based attack graphs.

\begin{table*}[htb!]
\centering
\caption{Annotated descriptions of each assessment aspect involved in the answer quality assessment process.}
\scalebox{0.88}{
\label{tab:annotation_rules}
\begin{tabular}{m{2cm}m{15.7cm}}
\toprule
\textbf{Aspects} & \textbf{Instructions} \\
\midrule
& 
\textbf{Score 1}: The description is entirely incorrect to the question and contains severely misleading information. \\
& \textbf{Score 2}: The description is partially correct but includes significant errors or irrelevant content. \\
\textbf{Accuracy} & \textbf{Score 3}: The description is mostly accurate but contains minor errors or ambiguous phrasing. \\
& \textbf{Score 4}: The description is accurate and correct but lacks direct image references. \\
& \textbf{Score 5}: The description is fully accurate, unambiguous, and directly supported by the image content. \\
\midrule
& 
\textbf{Score 1}: The description completely contradicts or is unrelated to the image information. \\
& \textbf{Score 2}: The description includes limited relevant details, but most content deviates from the image information. \\
\textbf{Consistency} & \textbf{Score 3}: The description partially aligns with the image information but contains irrelevant or redundant content. \\
& \textbf{Score 4}: The description closely adheres to the image information, with only minimal unrelated content. \\
& \textbf{Score 5}: The description is entirely based on the image information, with no extraneous or redundant elements. \\
\midrule
& 
\textbf{Score 1}: The description fails to address any critical aspects of the question, omitting all essential information. \\
& \textbf{Score 2}: The description addresses only a subset of the question, omitting most key details. \\
\textbf{Completeness} & \textbf{Score 3}: The description broadly covers the question’s requirements but lacks minor details. \\
& \textbf{Score 4}: The description comprehensively addresses the question, with only negligible omissions of minor details. \\
& \textbf{Score 5}: The description fully addresses all requirements of the question, providing thorough details. \\
\midrule
& 
\textbf{Score 1}: The description is entirely unrelated to cybersecurity and provides no value for threat analysis. \\
& \textbf{Score 2}: The description has marginal relevance, requiring substantial inference to connect to threat analysis. \\
\textbf{Relevance} & \textbf{Score 3}: The description partially relates to cybersecurity but lacks explicit ties to practical applications. \\
& \textbf{Score 4}: The description directly aligns with cybersecurity and offers moderate analytical value for threat analysis. \\
& \textbf{Score 5}: The description focuses heavily on cybersecurity and provides actionable insights into threat analysis.\\
\bottomrule
\end{tabular}
}
\end{table*}

\subsection{Brainstorming}

In multimodal threat intelligence analysis, accurately interpreting domain-specific threat images is essential for constructing attack graphs. These images visually depict cybersecurity scenarios and have different roles and contextual links in attack graphs.
Brainstorming phase mimics human cognition by analyzing image types and key features, generating critical questions to pinpoint image parsing priorities, thus underpinning follow-up work.

Specifically, in this phase, we first seed the LLMs with an initial set of questions, named leading questions~(Table~\ref{tab:ques-seed}). 
The leading questions contain a set of general questions that security practitioners are interested in for different types of threat images, which are grouped according to different image types. Secondly, the LLMs are guided by the prompts to generate a set of general questions for the current threat image based on the content of the image and the corresponding leading questions. Then, in order to better construct task associations with the attack graphs extracted from the textual modalities, the prompts guide the LLMs to generate a set of task-specific questions for the image based on the structure of the unimodal attack graph and the content of the current threat image. The general and task-specific question seeds are explained below.

\begin{itemize}[leftmargin=*, topsep=0.2pt,parsep=0pt]
      \item \textbf{General Question}. 
      This type of question focuses on mining information about the attributes that the images themselves have.
      Such as image subject, image type, and image source.
      The purpose is to prompt LLMs to perform a summary of the content of the image itself. 
      \item \textbf{Task-specific Question}. 
      This type of question focuses on the task-oriented nature of threat images in constructing attack graphs.
      Such as “\textit{What are the temporal features exhibited by the attack flow graph?}”
      The purpose of this is to motivate LLMs to mine threat image information from the attack graph construction task.
\end{itemize}

where the prompts used to generate the question pool are displayed in Table~\ref{tab:image-prompt}~(Question Generation).
Following the brainstorming phase, the general questions and task-based questions together form the question pool, which forms parsing guidance for the current threat intelligence images.

\subsection{Extraction}

The extraction phase builds on the brainstorming output. It uses a framework to help LLMs give accurate answers to parsing questions.  This approach facilitates the construction of attack graphs by introducing threat image context information as parsing support to create high quality answers.

Specifically, after generating the question pool, we will prompt the LLMs to answer the questions based on the question descriptions and threat image contents. In this phase, we focus on mining the content of threat images by answering the corresponding questions. 
In the actual threat image parsing process, we found that when LLM parses threat images without CTI contexts, the results are ambiguous and lack logical reliability. For example, the answer only describes abstract concepts but cannot identify specific threat objects. And such threat information is not suitable as a basis for the construction of multimodal attack graphs.
To address this issue, we enhanced the summarization module of AttacKG+ to extract two types of parsing-support information via image tag localization: image-aware context and global context. Section~\ref{5.5}~(Ablation Study) evaluates their importance for threat information extraction.

\begin{itemize}[leftmargin=*, topsep=0.2pt,parsep=0pt]
      \item \textbf{Image-Aware-Context}. In this setting, the Image-Aware-Context is derived from the summarization of the image context paragraph, such as image type and analysis approach, providing dynamic contextual information for threat images.
      \item \textbf{Global-Context}. In this setting, the Global-Context originates from the content abstract of the CTI report in which the threat image is located, encompassing the outline of the CTI report's subject matter and offering a macrosemantic framework for threat image parsing.
\end{itemize}

We control the output from three aspects: content limitation, topic relevance, and expression format through rule setting. The template of question answering prompts is shown in Table~\ref{tab:image-prompt}~(Question Answering).

\subsection{Verification-Question Filtering}

Question filtering aims to be more targeted and effective in parsing threat images, so as to exclude irrelevant questions.
This is because the quality of a question largely determines how helpful its answer is for attack graph construction.
The question filtering module consists of two phases: direct correlation question capture and answer-oriented question capture.

\begin{itemize}[leftmargin=*, topsep=0.2pt,parsep=0pt]
      \item \textbf{Direct correlation question capture}. 
      The idea of direct correlation question capture is to prompt LLMs to judge the quality of questions by taking the attack graph summary and domain rules as the basis.
      This method is suitable for cases where the direct formulation of the question has domain characteristics.
      For example, \textit{“What is the functional role of the malicious script in this image.”}
      In this case, “malicious script” has strong cyber security domain context.
      
      \item \textbf{Answer-oriented question capture}. 
      The idea of answer-oriented question capture is to first generate the answer to the corresponding question, and then judge the quality of the question based on the content of the answer and the hints of the LLMs.
      This method is suitable for cases where the direct formulation of the question does not have domain characteristics or the domain characteristics are weak.
      For example, \textit{“What trend does the graph in this image reflect?”}
      The description of the question does not have direct domain characteristics, but is based on the image identification and embedded text description of the chart.
      It is possible that threat information will be obtained that will assist in the construction of the attack graph.
\end{itemize}

After filtering the set of questions generated in the threat image parsing phase, we labeled the above two types of questions and fused them into the question candidate set for constructing multimodal attack graphs.

\subsection{Verification-Answer Refinement}
After completing question pool~(\textit{\textbf{Q}}) construction and question filtering, we input the questions along with the current threat images into the LLMs to obtain the corresponding set of question answers called answer pool~(\textit{\textbf{A}}) .
Specifically, these answers are our analysis of threat intelligence images from both general and task perspectives. However, we have to pay attention to two issues: 
First, it is necessary to determine that the questions currently generated are relevant to the attack graph construction rather than the minutiae. 
Second, we need to ensure that the answer to the question at hand is superior.


To address the above issues, we set up a self-learning module for LLMs to help optimize the quality of answer generation by iterative question-answering. 
Here, we propose two basic prompt optimization paradigms,  question-led~(Q-Led) and answer iteration~(A-Iteration), drawing on the idea of textual answer optimization.
Specifically, after completing the answer relevance analysis, we set up an answer assessment module to evaluate the quality of generated answers.
The quality criteria for the answers are divided into four levels, including: 
failing, satisfactory, good, excellent. 
As is shown in Table~\ref{tab:annotation_rules}, the assessment of answer quality is realized by guiding the LLMs to evaluate from four dimensions: accuracy, consistency, completeness, and relevance.
Accuracy represents whether the answer accurately answers the corresponding question.
Consistency represents whether the answer maintains content relevance to the threat image information.
Completeness represents whether the answer adequately answers the needs of the corresponding question.
Relevance represents whether the answer fits the cybersecurity domain in terms of presentation.

\begin{figure}[!t]
    \centering
    \includegraphics[width=0.65\linewidth]{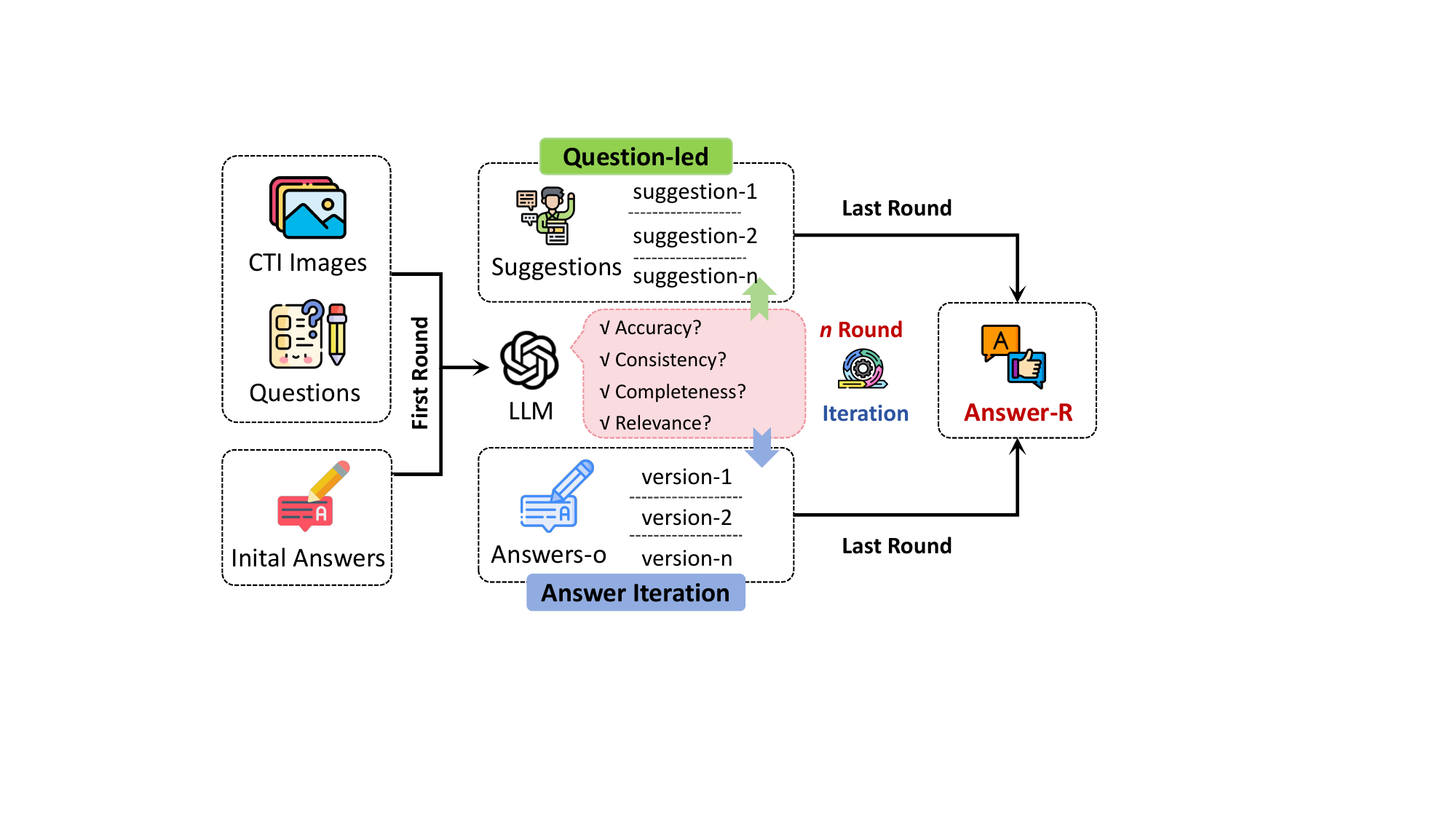}
    \caption{The threat image parsing answer optimization paradigm. \textbf{Question-led}, generate a target refinement parsing guide to guide the next round of question parsing of threat images. \textbf{Answer Iteration}, generate an initial answer, and refine the current answer in the next round based on the optimization comments. }
    \label{IA2}
\end{figure}

For answers below the good level, we perform answer refinement. 
The stopping condition for the number of iteration rounds of answer refinement is that (1) the current answer reaches "excellent/good" or (2) the set iteration threshold is reached. 
The optimization approach adopted in the answer refinement process is as follows:

\begin{equation}
   \mathrm{IA}^{\text {current}}=\text {LLM}\left(\mathrm{IA}^{\text {last}}+\text {Image}\right)
   \label{equation1}
\end{equation}

As shown in equation~\ref{equation1}, IA$^{\text {current}}$ denotes the threat image extraction answer in the current round, IA$^{\text {last}}$ denotes the threat extraction answer in the previous round, and Image denotes the current threat image file to be parsed. The idea of answer iteration is to evaluate the answers by co-inputting the generated answers and the belonging images into LLMs (without history). Then LLMs will output quality ratings and optimized comments for the current answer. Then the current answer is refined in the next iteration based on the optimization comments.

\begin{equation}
   \mathrm{IA}^{\text {current}}=\text {LLM}\left(\text {Suggestions}^{\text {last}}+\text {Image}\right)
   \label{equation2}
\end{equation}

In contrast, as shown in equation~\ref{equation2}. 
Suggestion$^{\text {current}}$ represents the optimization tips given by LLM in the previous round based on the answers and image content. The idea of question guidance is to evaluate the answers by co-entering the generated answers and the belonging images into LLMs (without history). 
Then LLMs will output quality ratings and the targeted refinement parsing guide for the current image.
It is worth noting that the suggestion differs from the answer comment. 
The suggestion is not oriented toward the previous round's answers but rather provides LLMs with an outline for a more comprehensive understanding of the image.
In the next round, the question, the image along the guide will be fed into LLMs together to get the new round of answers.

The processing of the above two answer refinement paradigms can effectively improve the quality of responses against cyber threat intelligence images.
The template of answer optimization prompts is shown in Table~\ref{tab:image-prompt}~(Answer Optimization).

\subsection{Integration}

According to the problem formulation in section~\ref{Problem Formulation}, this section presents a framework for constructing multimodal attack graphs based on LLMs. 
The construction of text-based attack graphs by AttacKG+ mainly involves (1) threat behavior extraction and (2) TTP labeling.
These two sub-tasks identify the behavior level and technique level threat information from the cyber threat intelligence respectively and portray the evolution flow of the whole threat event.
The details are as follows: for cyber threat intelligence, the method takes as input cyber threat intelligence in textual modality.
First, an attack graph ontology model is constructed via the STIX 2.0 standard~\cite{STIX}, and then threat atomic event~$(s, a, o, t)$ is extracted based on the ontology model, where~$t$ is derived from the relative timing of that atomic event in the intelligence.
Then, the dictionary of technology labels is constructed by the TTP matrix of MITRE ATT\&CK~\cite{Mitre}, identifying the technology labels corresponding to atomic events, and expanded to form the quintuple~$(s, a, o, t, p)$.
Next, based on the timestamp~$t$, the attack graph~$AG_t=\{(s, a, o, t, p)\}^N$ is formed.

It is worth considering that the image parsing module will generate multiple questions and their corresponding answers for the given image.
However, answers originating from the same threat image do not necessarily correspond to the same atomic event in the attack graph.
This is because the information in the images can complement different aspects of the threat event.
For example, the content in the attack flow diagram involves information about the flow of the entire current threat event.
To solve this problem, we construct each problem as a corresponding "threat enhancement reference".
Then, the structural enhancement of the attack graph is achieved by using the threat enhancement units as reference information.

First, we use LLMs to merge image topics, questions, and their corresponding answers to form threat enhancement reference extracted from the current threat image.
For example, ‘This is the temporal description (Question) of the protocol attack flowchart (Image Theme) as follows:(Answer)’.
In this case, the protocol attack flowchart is the topic of the current threat image, the temporal description is the corresponding particular question, and the specific content is the answer obtained through three stages of brainstorming, extraction, and verification.
Then, we use the threat enhancement information as a basis to correlate and enhance the current attack graph.
We regulate three kinds of functional enhancements of threat image attack graphs:

\begin{itemize}[leftmargin=*, topsep=0.2pt,parsep=0pt]
      \item \textbf{Node Extension}. 
      Adding new nodes or checking the node descriptions of existing attack graphs to supplement data or attack resources through threat image information.
      \item \textbf{Relation Update}. 
       If the image reveals a new attack action (such as \textit{deliver} or \textit{execute}), add or replace the relation and description. 
       \item \textbf{Technique Addition}. 
       Match the added technique (such as \textit{T1204.002-User Execution}) according to the MITRE ATT\&CK framework. 
\end{itemize}

Through the targeted parsing and fusion of threat images by LLMs, the attack graphs combined with threat images can be greatly improved in the three levels of data integrity, structural flow and technical richness.

\section{Evaluation} 
\label{Evaluation}

We conduct extensive experiments to answer the following research questions:


\begin{itemize}[leftmargin=*, topsep=0.2pt,parsep=0pt]
    \item \textbf{RQ1:} 
    How does MM-AttacKG perform against existing threat information extractors?~(see Section \ref{5.2})
    \item \textbf{RQ2:} 
    How effective is MM-AttacKG in enhancing attack graph with threat image information?~(see Section \ref{5.3})
    \item \textbf{RQ3:} 
    Whether each key module in MM-AttacKG effective?~(see Section \ref{5.4}) 
    \item \textbf{RQ4:} 
    How does each technical module function within the MM-AttacKG framework?~(see Section \ref{5.5}) 
\end{itemize} 


\subsection{Experiments Setting}
To ensure the rationality and reproducibility of the evaluation phase for MM-AttacKG, we describe the experiments setting, including the dataset, baseline methods, and implementation details.

\subsubsection{Dataset}
To evaluate MM-AttacKG, we constructed the dataset by collecting cyber threat intelligence reports from Cisco Talos Intelligence Group~\cite{Cisco}, Microsoft Security Intelligence Center~\cite{Microsoft}.
We utilize AMinerU-PDFScanner~\footnote[2]{\url{https://github.com/liuhuapiaoyuan/MinerU-PDFScanner}} to extract each CTI report into two parts: textual content and threat images. For the textual content, we extract it into attack graphs using the construction process provided in AttaKG+~\cite{AttacKG+}, and manually verify the extraction results. For the threat images, we follow the following process: first, due to the layout, the original CTI reports or web pages often contain a large number of irrelevant, low-quality threat images. 
Therefore, we preliminarily filter the threat images and remove those with the following problems.
(a) containing irrelevant information such as logos, advertisements, etc. (b) presence of occlusion or strong watermarks; (c) weakly informative visual samples; (d) poor image quality or lack of clarity; and (e) crippled images or incomplete graphics.
Then, to achieve this, we set up image rule filtering rules, write the prompting scheme, and use LLMs to analyze the content and board style of images in order to eliminate irrelevant images. Meanwhile, we arranged researchers with cybersecurity background to proofread again.
Finally, we integrate textual content and threat images. 
Then name the dataset as \textbf{A}ttack \textbf{G}raph-\textbf{LLM}-\textbf{m}ulti\textbf{m}odal, short for AG-LLM-mm.
Three postgraduate students from our team acted as participants in the manual assessment of the CTI reports. Participants exchanged views after completing the assessments individually, which was secondarily discussed in order to obtain a comprehensive assessment.

\begin{table*}[htpb]
  \renewcommand\arraystretch{1.5}
  \centering
  \caption{Effectiveness of MM-AttacKG for multimodal threat information extraction. For each column, the bold number indicates the best performance, and the underlined number corresponds to the second-best performance. Human Anotation-Text represents the entity, relation and technique extraction of threat intelligence text under the manual annotation process.}
  \label{table-R1-R2}
  \small
  {%
    \renewcommand{\arraystretch}{1.3}
    \setlength{\tabcolsep}{8pt}
    \setlength{\arrayrulewidth}{0.4pt}
    \setlength{\dashlinedash}{1pt}
    \setlength{\dashlinegap}{2pt}
    \scalebox{0.75}{%
    \begin{tabular}{>{\centering\arraybackslash}p{3.5cm}|>{\centering\arraybackslash}p{1.4cm}>{\centering\arraybackslash}p{1.4cm}>{\centering\arraybackslash}p{1.4cm}|>{\centering\arraybackslash}p{1.4cm}>{\centering\arraybackslash}p{1.4cm}>{\centering\arraybackslash}p{1.4cm}|>{\centering\arraybackslash}p{1.4cm}>{\centering\arraybackslash}p{1.4cm}>{\centering\arraybackslash}p{1.4cm}}
      \hline
      \multirow{2}{*}{Method}
        & \multicolumn{3}{c|}{Entity}
        & \multicolumn{3}{c|}{Relation}
        & \multicolumn{3}{c}{Technique} \\
        & Precision & Recall & F-1
        & Precision & Recall & F-1
        & Precision & Recall & F-1 \\
      \hline
      \multicolumn{10}{c}{\itshape Text-based Method} \\
      \hline
      Extractor
    & 0.6568 & 0.5387 & 0.5919 & 0.2158 & 0.1026 & 0.1391 & - & - & - \\
      AttacKG
    & 0.5580 & 0.2612 & 0.3559 & - & - & - & 0.2060 & 0.3399 & 0.2565 \\
      AttacKG+
    & \textbf{0.7701} & 0.5294 & 0.6274 & \textbf{0.7693} & 0.6806 & 0.7222 & 0.4502 & 0.4481 & 0.4491 \\
     \hline
      Human Anotation-Text & 1.0000 & 0.4559 & 0.6263 & 1.0000 & 0.6820 & 0.8109 & 1.0000 & 0.6547 & 0.7913 \\
      \hline
      \multicolumn{10}{c}{\itshape Image-enhanced Method} \\
      \cdashline{1-10}[4pt/1pt]
      ICL
    & 0.6901 & 0.7326 & \underline{0.7107}& 0.7106 & 0.8261 & \underline{0.7640}& 0.4948 & 0.5383 & 0.5156 \\
      CoT
    & 0.6805 & \underline{0.7432}& 0.7105 & 0.6949 & \underline{0.8383}& 0.7599 & \underline{0.5063}& \underline{0.5508}& \underline{0.5277}\\
      MM-AttacKG
    & \underline{0.7224}& \textbf{0.8280} & \textbf{0.7716} & \underline{0.7460}& \textbf{0.8973} & \textbf{0.8147} & \textbf{0.5256} & \textbf{0.6232} & \textbf{0.5703} \\
      \hline
    \end{tabular}
    }%
  }%
\end{table*}

\subsubsection{Baseline Methods}
To assess the effectiveness of MM-AttacKG for threat information extraction and image-enhanced attack graph integration, we compare it with four text-based threat information extractors~(e.g.,  AttacKG~\cite{AttacKG}, EXTROCTOR~\cite{EXTRACTOR}, AttacKG+~\cite{AttacKG+}).
Meanwhile, we migrate two LLM-based methods~(ICL~\cite{dong2024surveyincontextlearning} and CoT~\cite{wei2023chainofthoughtpromptingelicitsreasoning}) to the multimodal threat information extraction and attack graph aggregation task.
More details are in Appendix Table~\ref{tab:image-prompt-ICL} and Table~\ref{tab:image-prompt-CoT}.

\subsubsection{Implementation Details}
As the field of cybersecurity requires transparent information processing and prevents information leakage, commercial API services may be unstable and face service restrictions.
We have chosen to use the open-source multimodal large language model from the Qwen~\cite{EXTRACTOR} series~(e.g., Qwen2.5-VL-72B, Qwen2.5-VL-32B, Qwen2.5-VL-7B) for relevant experiments.
We hosted the large language models on an AliCloud Elastic Compute Service~(ECS)~\footnote[3]{\url{https://www.aliyun.com/}}.  instance and developed all implementation modules within a Python 3.10.14 environment.
To ensure the reproducibility of the research, we fixed the temperature parameter of the proprietary LLMs used to 0.7 and set the seed parameter to a constant value.
Meanwhile, to keep the experiments manageable, for different LLMs prompt methods, we provide them with exactly the same context and specify output limits in the same format.
ethod, we turn on the relevant optimization setting to ensure the construction quality of the attack graph.
To prevent invalid responses, we limit the maximum output length to 512 tokens when parsing threat images.


\subsection{Performance Comparison for Threat Extraction~(RQ1)}
\label{5.2}

This experiment aims to quantify the effectiveness of images in enhancing CTI understanding by counting the threat information that can be mined from CTI reports. Specifically, we analyze the performance improvement of threat information extraction with the introduction of visual information.

Based on the problem formulation in section~\ref{Problem Formulation}, a cyber threat event consists of multiple threat atomic events in temporal order. 
These atomic events are represented in the threat behavior graph as interconnected quintets $(s, a, o, t, p)$, where $s$ and $o$ are coordinated for entity extraction, and actions $a$ and entity-entity relations are coordinated for relation extraction. 
$p$ represents the TTP technique label involved in the current behavior.
By parsing the threat image information, more information can be gained to introduce to the attack graph.
The gained information includes three aspects:
(1) Entity. Subjects and objects in cyber threat behavior.
(2) Relation. Descriptions of correlations between subjects and objects in cyber threat behavior.
(3) Technique. Mining technique labels from threat images based on the TTP matrix.
Table~\ref{table-R1-R2} summarizes the effectiveness of MM-AttacKG for multimodal threat information extraction.
Our findings are: 

(1) Advantages of multimodal integration. The image-enhanced method demonstrates superior performance across entity extraction, relation extraction, and technical identification tasks compared to Text-based methods. This indicates that incorporating visual information significantly improves threat intelligence mining efficacy and enhances perception of complex threat scenarios.

(2) Image-enhancement methods comparison. Among multimodal approaches, the MM-AttacKG model achieves superior F-1 scores across all three subtasks relative to ICL and CoT. This suggests that MM-AttacKG’s architecture provides stronger multimodal integration capabilities and semantic comprehension, offering more reliable foundations for cybersecurity threat intelligence analysis.

(3) Subtask difficulty disparities. Technical identification exhibits markedly lower F-1 scores (e.g., 0.4491 for AttacKG+ in technical identification vs. 0.6274 and 0.7222 in entity/relation extraction). This highlights the heightened complexity of technical identification tasks and underscores the need for algorithmic improvements to enhance semantic parsing capabilities in this domain.

(4) Limitations of manual annotation. Though manual text annotation by humans yields high precision, it is constrained by the absence of image information, resulting in low recall rates. Additionally, this process is labor-intensive. The dependence on domain experts further limits scalability and real-time applicability. These factors underscore the critical need for automated methods in large-scale data processing and continuous monitoring within cybersecurity contexts.

\begin{figure*}[!t]
    \centering
    \includegraphics[width=1\linewidth]{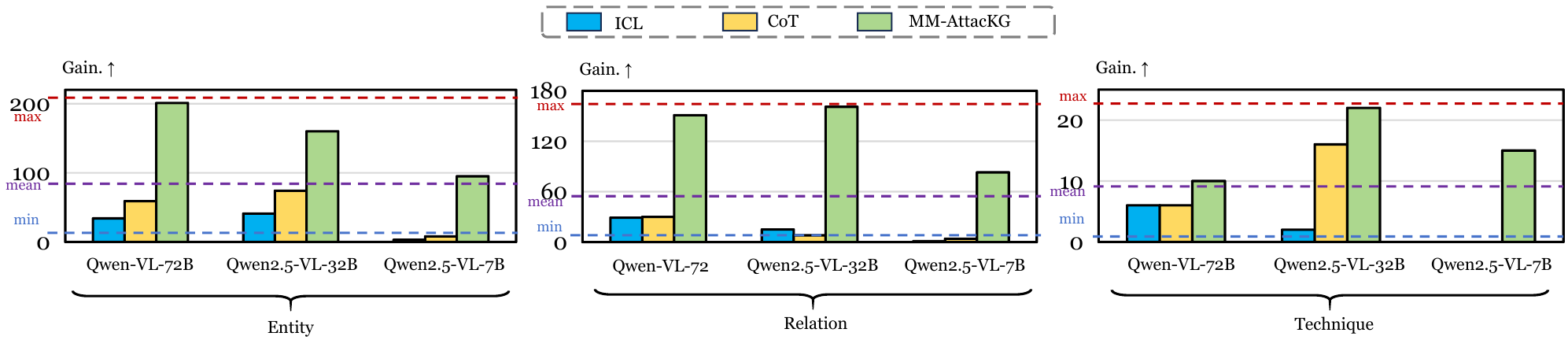}
    \caption{Performance of MM-AttacKG's threat gain in different prompting schemes and LLM versions.
}
    \label{Position}
\end{figure*}

\subsection{Performance of Attack Graph Integration~(RQ2)}
\label{5.3}

To address RQ2, we introduced In-Context Prompt Learning (ICL) and Chain of Thought (CoT) as comparative methods for constructing multimodal attack graphs.   ICL enables LLMs to learn directly from input examples or instructions to complete tasks.   CoT allows learners to solve problems through step-by-step reasoning, mimicking the human thought process for complex problem-solving.   MM-AttacKG enhances the performance of attack graph construction by integrating prompts from three phases: brainstorming, extraction, and verification.

For the dataset in the experimental setup, we implemented the above prompt schemes in three versions of Qwen-VL. We evaluated the performance of the attack graphs in terms of entity, relationship, and the gain from utilizing image modality information. Namely, the incremental information of the updated multimodal attack graph over the text-based attack graph at the entity, relation, and technology levels. The evaluation results are shown in Figure~\ref{Position}.
The experimental results indicate that: (1) Across all versions of large language models, MM-AttacKG outperforms the other two prompt schemes in terms of information gain in the entity, relationship, and technique dimensions. (2) We observed the most significant performance fluctuations in ICL across different LLM versions, suggesting that merely providing operational examples results in poor robustness of LLMs for CTI image parsing and attack graph construction.   Although the CoT prompt scheme demonstrated strong robustness in constructing mult-modal attack graphs across various LLM versions, its chain-like thinking process lacks an active guidance mechanism for performance enhancement, resulting in limited information gain.   In summary, MM-AttacKG is superior as a multimodal attack graph construction scheme.

\begin{table*}[!t]
  \renewcommand\arraystretch{1.5}
  \centering
  \caption{Ablation study. For each column, the bold number indicates the best performance, and the underlined number corresponds to the second-best performance. \textbf{Support Context for Image Threat Information Extraction}: I stands for Image-Aware-Context, G stands for Global-Context. \textbf{Image Threat Information Extraction Module}: B stands for Brainstorming, V stands for Verification.}
  \label{table3}
  \small
  {%
    \renewcommand{\arraystretch}{1.3}
    \setlength{\tabcolsep}{8pt}
    \setlength{\arrayrulewidth}{0.4pt}
    \setlength{\dashlinedash}{1pt}
    \setlength{\dashlinegap}{2pt}
    \scalebox{0.78}{%
    \begin{tabular}{>{\centering\arraybackslash}p{3.5cm}|>{\centering\arraybackslash}p{1.3cm}>{\centering\arraybackslash}p{1.3cm}>{\centering\arraybackslash}p{1.3cm}|>{\centering\arraybackslash}p{1.3cm}>{\centering\arraybackslash}p{1.3cm}>{\centering\arraybackslash}p{1.3cm}|>{\centering\arraybackslash}p{1.3cm}>{\centering\arraybackslash}p{1.3cm}>{\centering\arraybackslash}p{1.3cm}}
      \hline
      \multirow{2}{*}{Method}
        & \multicolumn{3}{c|}{Entity}
        & \multicolumn{3}{c|}{Relation}
        & \multicolumn{3}{c}{Technique} \\
        & Precision & Recall & F-1
        & Precision & Recall & F-1
        & Precision & Recall & F-1 \\
      \hline
      \multicolumn{10}{c}{\itshape Our Method} \\
      \cdashline{1-10}[4pt/1pt]
      MM-AttacKG
    & 0.7224 & \textbf{0.8280} & \textbf{0.7716} & 0.7460 & \textbf{0.8973} & \textbf{0.8147} & \textbf{0.5256} & \textbf{0.6232} & \textbf{0.5703} \\ \hline
    \multicolumn{10}{c}{\itshape Support Context for Image Threat Information Extraction
} \\
      \cdashline{1-10}[4pt/1pt]
w/o I\&G& 0.6413 & 0.7759 & 0.7022 & 0.7298 & 0.8583 & 0.7888 & 0.4543 & 0.5795 & 0.5093 \\
      w/o G& 0.6449 & 0.7913 & 0.7106 & 0.7116 & 0.8581 & 0.7780 & 0.4876 & 0.6037 & 0.5395 \\
      w/o I& 0.6768 & 0.7740 & 0.7221 & 0.7027 & 0.8539 & 0.7710 & 0.4697 & 0.5826 & 0.5201 \\
      \hline
      \multicolumn{10}{c}{\itshape Image Threat Information Extraction
Module
} \\
      \cdashline{1-10}[4pt/1pt]
w/o B\&V& \textbf{0.7710} & 0.6000 & 0.6749 & \textbf{0.7887} & 0.7947 & 0.7917 & \underline{0.5005}& 0.5829 & 0.5386 \\
      w/o V& 0.7086 & \underline{0.8247}& \underline{0.7623}& 0.7386 & \underline{0.8963}& \underline{0.8099}& 0.4859 & \underline{0.6225}& \underline{0.5458}\\
      w/o B& \underline{0.7607}& 0.6028 & 0.6726 & \underline{0.7862}& 0.7936 & 0.7899 & 0.4820 & 0.5795 & 0.5263 \\
      \hline
    \end{tabular}
    }%
  }%
\end{table*}


\subsection{Ablation Study~(RQ3)}
\label{5.4}
In our research, to evaluate the influence of support context and various components on the performance of MM-AttacKG, we carried out two types of ablation studies through the controlled removal of support context and key modules.
To confirm the facilitating role of supporting information in threat image information extraction, we removed specific configurations, namely Image-Aware-Context\&Global-Context, Image-Aware-Context, and Global-Context.
In order to verify the importance of key components for threat image information extraction, we removed the settings of Brainstorming\&Verification, Brainstorming, and Verification.
The outcomes of these ablation experiments are presented in Table~\ref{table3}.

(1) Multimodal Fusion Advantage. MM-AttacKG, through integrating Image-Aware-Context and Global-Context with component collaboration (Brainstorming, Extraction, Verification), significantly outperforms text-only variants in entity recognition and relation extraction. This highlights the effectiveness of multimodal fusion in enhancing threat information identification, making it a key driver of model performance.

(2) Synergy of Textual Support. Image-Aware-Context and Global-Context support each excel in different tasks. Image-Aware-Context support boosts relation extraction recall with dynamic context, while Global-Context support improves entity recognition precision with macro semantic frameworks. It is their synergy that powerfully boosts the model's performance in a multitude of tasks.

(3) Module Collaboration Superiority. Comparing variants without verification module and those without both brainstorming and verification module shows that removing only verification improves entity recognition but harms threat information extraction. The full MM-AttacKG model achieves better balanced and slightly superior performance in all tasks, proving its component configuration effectively meets diverse task needs.

(4) Overall Performance and Design Rationality. MM-AttacKG delivers consistently strong performance across entity recognition, relation extraction, and threat information extraction. This validates its effectiveness and robustness in multi-task scenarios.

\subsection{In-depth and Analysis of Key Modules~(RQ4)}
\label{5.5}

Given that our method is an attack graph construction pipeline composed of multiple modules, to address RQ4, we divide this question into three aspects: the diversity of questions generated by the brainstorming module, the necessity of the question filtering module, and the superiority of the answer refinement module.
For the brainstorming module, we focus on whether the set of questions generated based on leading questions can sufficiently cover more main points of mining CTI images and whether there is enough distinction among the questions.
For the question filtering module, we focus on the necessity of question filtering and how much unnecessary attention information in CTI images we can reduce through this method.
For the superiority of the answer refinement module, we focus on the changes in the distribution of answers of different quality levels with the increase of iterative rounds. Then, analyzing the differences between the two optimization paradigms.

\begin{figure}[!t]
    \centering
    \includegraphics[width=0.45\linewidth]{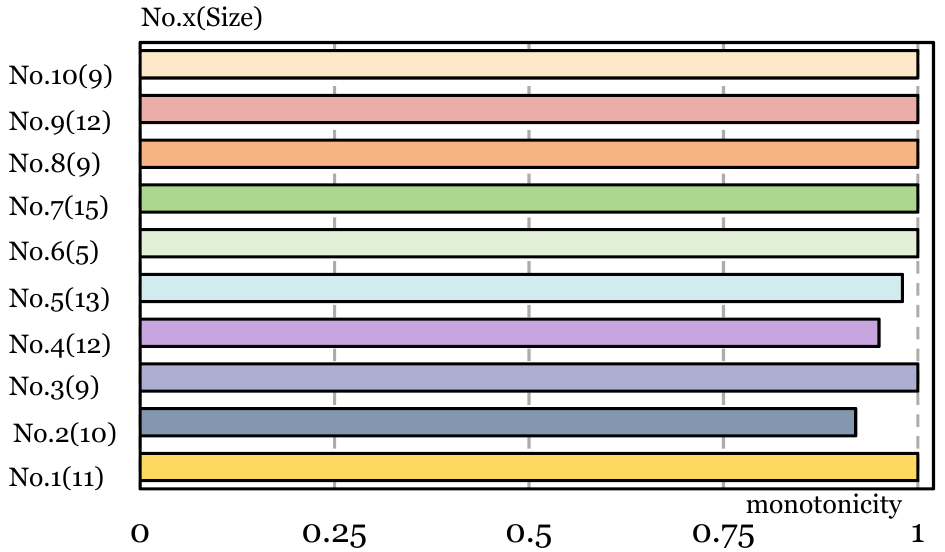}
    \caption{
    Brainstorming questions generate evaluation situations. The abscissa represents the monotonicity of the question set, and the ordinate represents the scale of the current image question set.
}
    \label{Monotonicity}
\end{figure}

\subsubsection{The Diversity of Questions Generated By Brainstorming} 
During the brainstorming phase, two types of questions, namely general questions and task-specific questions were generated.
On average, 21 questions were generated per CTI image, a quantity comparable to that in manual judgment scenarios. Specific question examples are provided in Table~\ref{tab:ques-seed}.
To assess question diversity, we adopted the monotonicity indicator from work~\cite{e26121075}. 
Here, monotonicity is evaluated by measuring the similarity among questions targeting the same threat image. Higher monotonicity indicates more similar questions and thus poorer diversity.
We analyzed the monotonicity of question sets for a random sample of 10 CTI images from the dataset, with results shown in Figure~\ref{Monotonicity}. 
We found that (1) the number of questions varies with the CTI image, reflecting that the information richness of the CTI image affects the number of questions generated.Richer information leads to more threat information being mined. 
(2) The monotonicity of question sets remains stable and close to [1], indicating low similarity among the generated questions, thus demonstrating high diversity.

\begin{table}[htp]
   \renewcommand\arraystretch{1.5}
   \centering
   \caption{Distribution of question types in CTI images.}
   \label{table-R3} 
   \resizebox{0.55\textwidth}{!}{
      \begin{tabular}{l|ccc}
         \hline
         \textbf{Scheme} & \textbf{Direct correlation} & \textbf{Answer-oriented}  & \textbf{Non-Related} \\ \hline
         Proportion                & 0.4699& 0.4185& 0.1116\\
         Size                      & 602& 536& 143\\ \hline
      \end{tabular}
   }
\end{table}

\subsubsection{The Necessity of Question Filtering Module}
The question filtering module aims to capture the set of important questions from two aspects: direct correlation and answer-oriented, thereby enhancing the overall quality of the questions generated in the brainstorming phase through filtering.   During the construction of AG-LLM-mm, 1,281 questions were initially generated.   However, some of these questions had unclear wording or low relevance.   After question filtering, the overall question distribution is shown in Table~\ref{table-R3}.
It can be observed that:
(1) the proportion of direct correlation questions is 0.4699, and the proportion of answer-oriented questions is 0.4185. This indicates that solely using question direct expression strategies to filter relevant questions will miss many valuable aspects of CTI images. 
(2) Non-Related questions account for 143, which shows that the initial set of questions generated in the brainstorming phase cannot be guaranteed to fully meet the requirements of attack graph construction.   
Therefore, further filtering of questions is necessary.

\begin{figure*}[!t]
    \centering
    \includegraphics[width=0.80\linewidth]{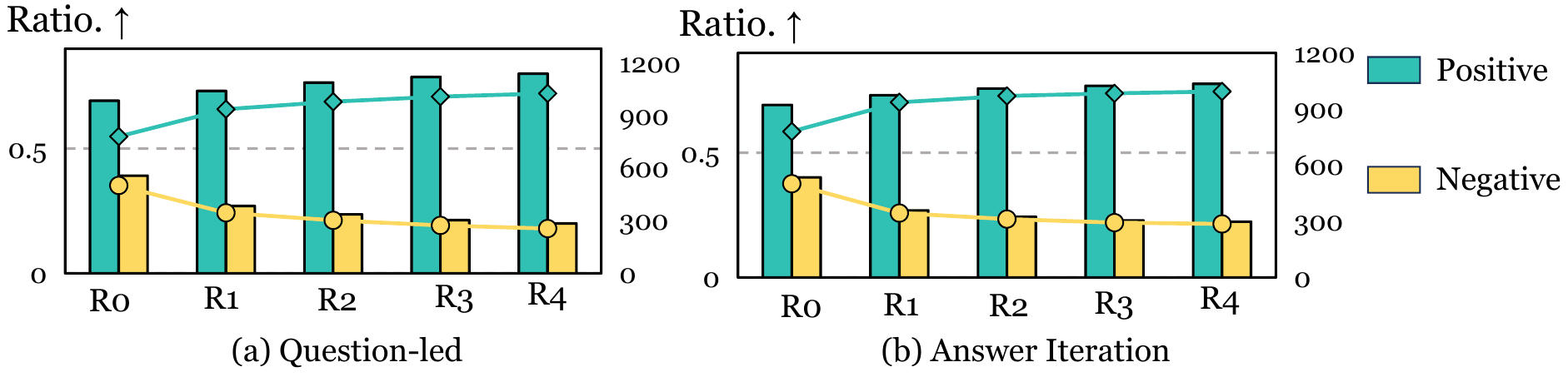}
    \caption{The answer quality distribution changes with the number of refinement rounds.
     (1) \textbf{Positive}: Answers scored as excellent or good are marked as positive. 
     (2) \textbf{Negative}: Answers scored as failing or satisfactory are marked as negative.
}
    \label{Answer_Refined}
\end{figure*}

\subsubsection{The Superiority of Answer Refinement Module}
The answer refinement module refines the parsing results of CTI images through multiple rounds based on optimization paradigms to enhance the quality of answers. We propose two optimization paradigms: question-led and answer iteration. question-led focuses on using the current answer as a case study to suggest optimizations for the next round of answers, while answer iteration emphasizes supplementing and optimizing the current answer.
To evaluate the effectiveness of the answer refinement module in improving answer quality, we designed an experiment as shown in Figure~\ref{Answer_Refined}, setting the number of optimization rounds from 1 to 4 and tracking the changes in answer quality with each round.
Here we set the answer to be marked as positive when the score is excellent or good, indicating that it meets the quality requirements, and negative (failing/satisfactory) when it does not.
The results show that: (1) As the number of optimization rounds increases, the overall quality rating of the answers consistently improves, demonstrating the effectiveness of the answer refinement module in enhancing answer quality. (2) The question-led optimization paradigm converges more easily, indicating that suggesting new answers based on recommendations may be more effective in addressing the questions than refining the current answers.
(3) As the number of optimization rounds increases, the answer refinement effect gradually decreases, which means gradually approaching the optimization limit.


 



\begin{figure*}[htp]
    \centering
    \includegraphics[width=0.90\linewidth]{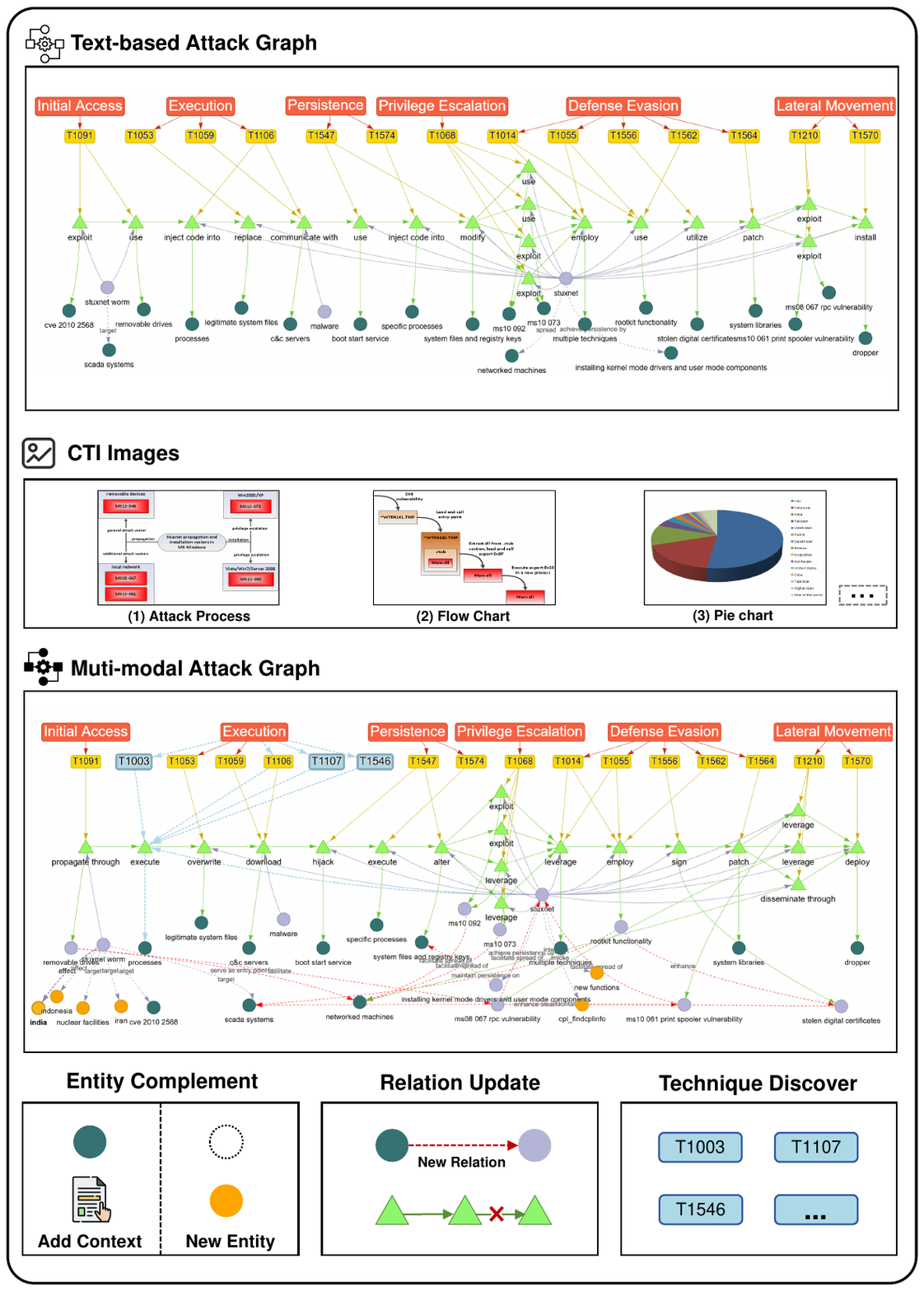}
    \caption{Example of multimodal attack graph Constrction~(MM-AttacKG).}
    \label{case_study}
\end{figure*}

\subsection{Case Study}
\label{sec:case study}

This section delves into the efficacy of MM-AttacKG in real-world security tasks through case studies. In cybersecurity, professionals and researchers often conduct in-depth analyses of existing threats based on CTI reports. This process focuses on two key aspects: one is to accurately extract core entities linked to the attack process from intricate event info and remove irrelevant redundancy; the other is to bridge the inherent knowledge gap between CTI reports and actual attacks, thus precisely determining tactical stages and identifying attack techniques. This analysis heavily depends on the comprehensive evaluation and information extraction of textual and image information.

MM-AttacKG, with its unique advantages, leverages structured knowledge for in-depth event analysis, effectively meeting the above two goals. As shown in Figure \ref{case_study}, based on MM-AttacKG's advanced concept, firstly, it interprets threat texts to structurally describe the attack process or event, accurately building a text-based attack graph. Then, it parses and extracts threat image info, seamlessly integrating it with the text-based attack graph to innovatively form a multimodal attack graph.
We've created an intuitive and efficient visualization interface for MM-AttacKG using pyvis~\footnote[4]{\url{https://github.com/WestHealth/pyvis}}. 

Take the Stuxnet worm attack as an example. MM-AttacKG first accurately extracts structured knowledge of the threat event scenario from the text description. It can clearly identify the 6 key tactical stages and 14 specific techniques in the attack, like using \textit{T1091-Replication Through Removable Media} for initial access, \textit{T1574-Hijack Execution Flow} for execution, and \textit{T1055-Process Injection} for defense evasion. Next, based on the techniques and entities related to the threat event, MM-AttacKG accurately infers the complex system environment of the intrusion activity. Specifically, the attack graph details how the Stuxnet worm gains initial access via malicious .LNK files and vulnerabilities, achieves precise system control and persistence through the main module, and escalates privileges using a zero-day vulnerability. It can scan network shares for lateral movement and employs various sophisticated techniques to evade detection. Finally, it communicated with the C\&C server through the encrypted channel to receive updates and instructions, and adopted threat behaviors to realize serious interference with the normal operation of industrial equipment.

Notably, after integrating threat image information, MM-AttacKG captures three new techniques (\textit{T1003-OS Credential Dumping}, \textit{T1107-Function hooking}, \textit{T1546-Event Triggered Execution}) in the report, derived from deep information extraction of corresponding threat images. Meanwhile, MM-AttacKG offers a more detailed entity set, uncovers deeper and more hidden threat relationships, and accurately supplements and improves the attack process. 
This enriches and optimizes the attack graph's description of threat events, successfully integrating text and image into an advanced multimodal attack graph.
In conclusion, MM-AttacKG provides a solid and reliable foundation of key information to reconstruct threat events with its superior performance, showing great potential for application and value in cybersecurity analysis.


\section{Conclusion and Future Work}
\subsection{Conclusion}

In this work, we introduces CTI images into attack graph construction for the first time by analyzing the role of image information in the cyber threat intelligence analysis process.

Leveraging the superior multimodal information understanding capabilities of LLMs, we propose an automated LLM-based framework (MM-AttacKG) for constructing multimodal attack graphs. Given the performance advantages of MLLMs and the parsing requirements of CTI images, we design a multistage prompt scheme that integrates brainstorming, extraction, verification and integration. As a byproduct, we construct the multimodal threat intelligence dataset AG-LLM-mm. Finally, through detailed experiments, we demonstrate that incorporating CTI images enhances the overall performance of attack graphs and that LLMs hold great potential for multimodal attack graph construction.

\subsection{Future Work}
In this work, we present a novel multimodal attack graph construction method and develop an automated image modality information extraction pipeline using LLMs. Although MM-AttacKG is novel and effective, there are still limitations and areas for further research in CTI parsing.

 \textbf{Further Integration of Multimodal Information}. 
Current analysis processes for multimodal threat information separate and individually understand different modalities, but the parsing requirements of different modalities may be interdependent.  Therefore, constructing an end-to-end workflow to simultaneously analyze different modalities of threat information and perform cross-verification and support can lead to a deeper understanding of CTI.

 \textbf{Joint Analysis of Multi-Source Threat Reports}. 
There may be event correlations between multiple threat actors, such as different application scenarios and targets of the same malware.  By tracking the version iterations of the malware, one can infer the ongoing confrontation between defense strategies and attackers, as well as potential future victims.  Analyzing multi-source threat intelligence to build joint analysis strategies can provide deeper insights into the evolution of threats.

\textbf{Training Domain-Specific Cybersecurity LLMs}.
General-purpose large models still have gaps in handling complex cybersecurity problems.  Therefore, exploring how to pre-train LLMs with domain-specific data and perform targeted optimizations is valuable and can reduce processing costs.

\section*{Authorship contribution statement}
Yongheng Zhang: Writing original draft, Software, Methodology, Data curation, Conceptualization. Xinyun Zhao: Writing review \& editing, Supervision, Formal analysis, Conceptualization. Yunshan Ma, Haokai Ma, Yingxiao Guan: Writing review \& editing, Visualization, Investigation, Data curation. Guozheng Yang, Yuliang Lu, Xiang Wang: Writing review \& editing, Supervision.

\section*{Declaration of competing interest}
The authors declare that they have no known competing financial interests or personal relationships that could have appeared to influence the work reported in this paper.

\section*{Data availability}
The code and the corresponding dataset will be released upon acceptance.

\section*{Funding}
The project was supported by Open Fund of Anhui Province Key Laboratory of Cyberspace Security Situation Awareness and Evaluation.

\bibliographystyle{unsrt}  
\bibliography{references}


\appendix
\section{}
\label{AA}
\begin{figure*}[!t]
    \centering
    \includegraphics[width=0.95\linewidth]{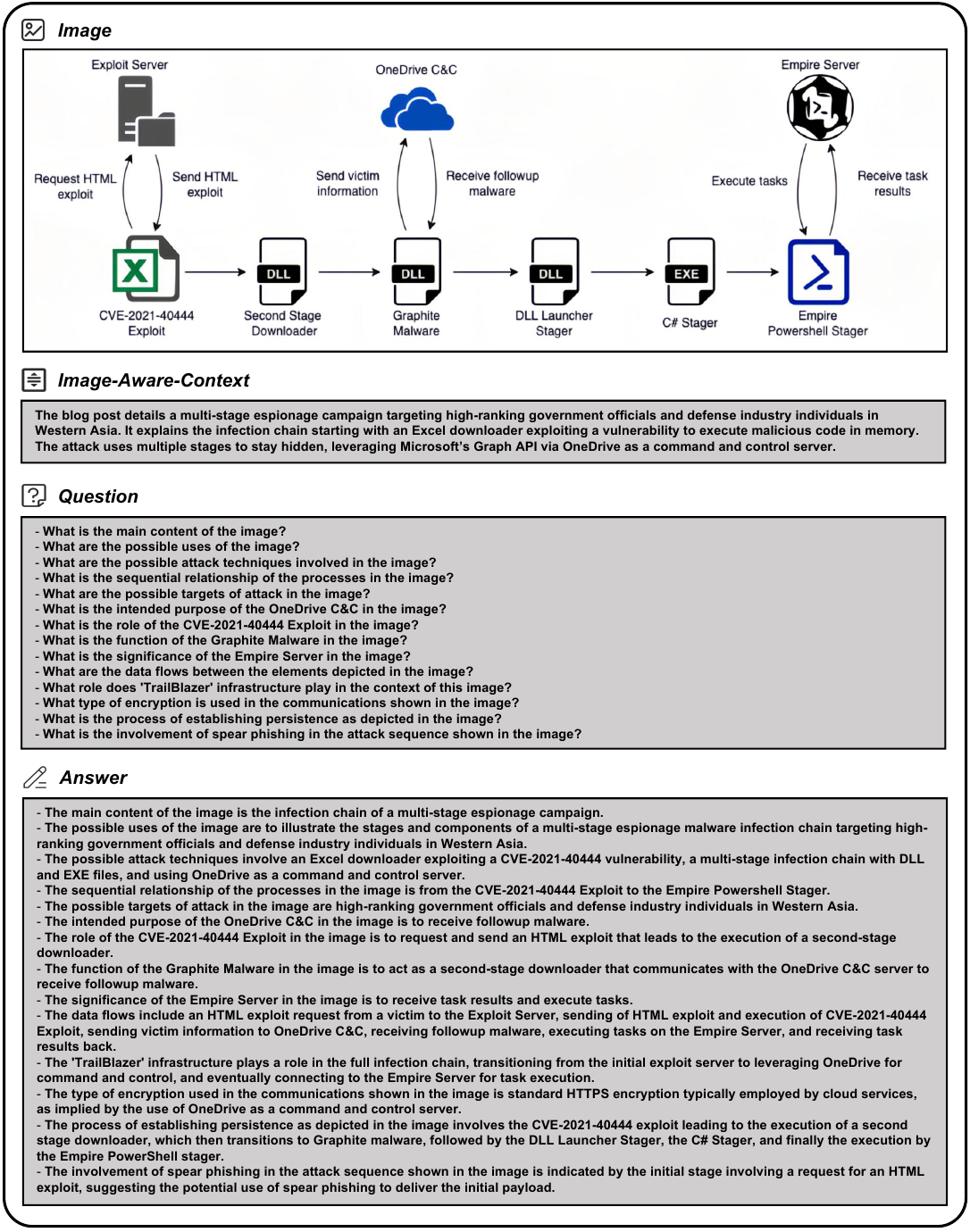}
    \caption{QA Example}
    \label{QA_Example}
\end{figure*}

\begin{figure*}[!t]
    \centering
    \includegraphics[width=0.95\linewidth]{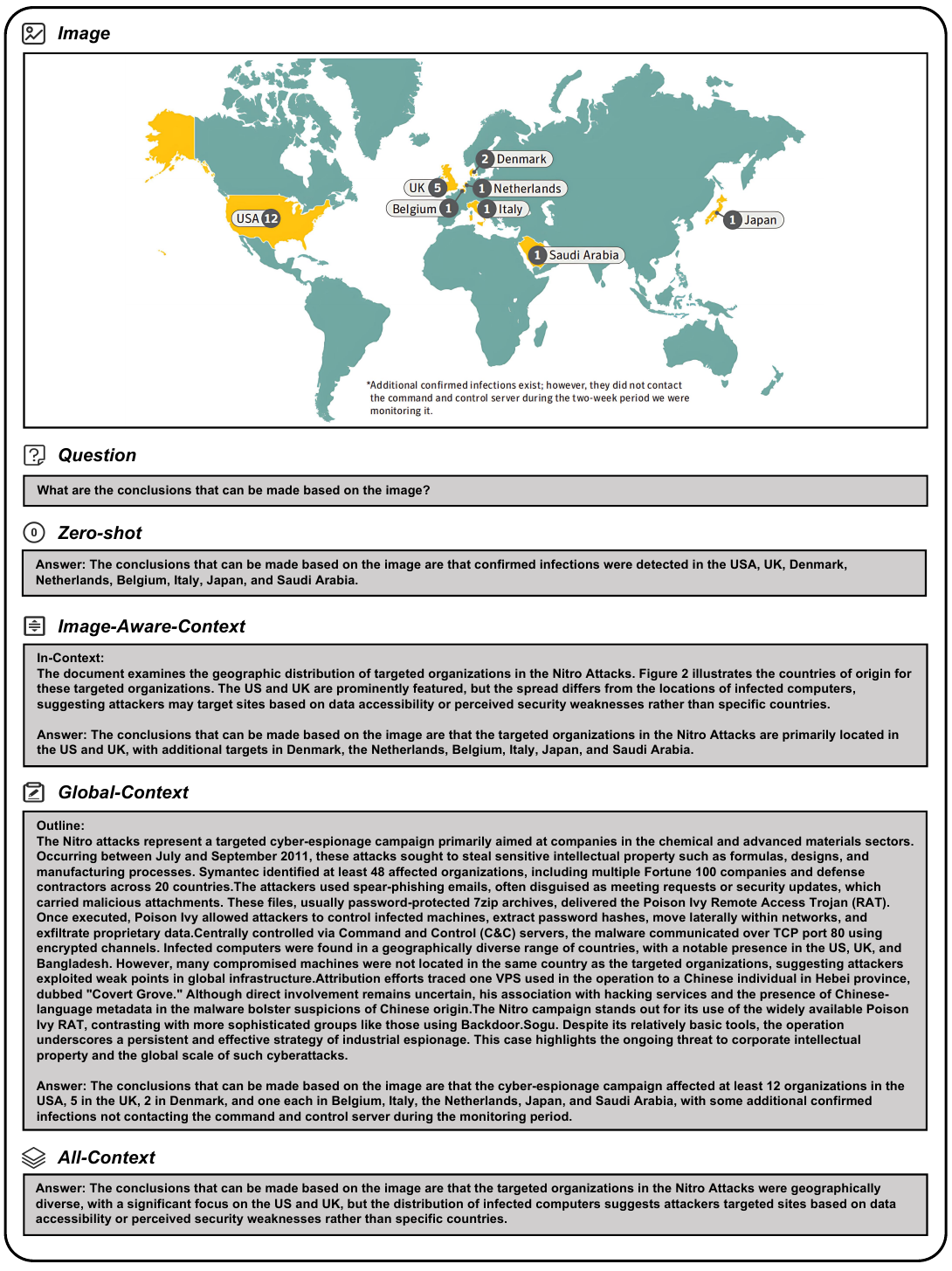}
    \caption{Answer Example}
    \label{AnswerExample}
\end{figure*}

\begin{figure*}[!t]
    \centering
    \includegraphics[width=0.95\linewidth]{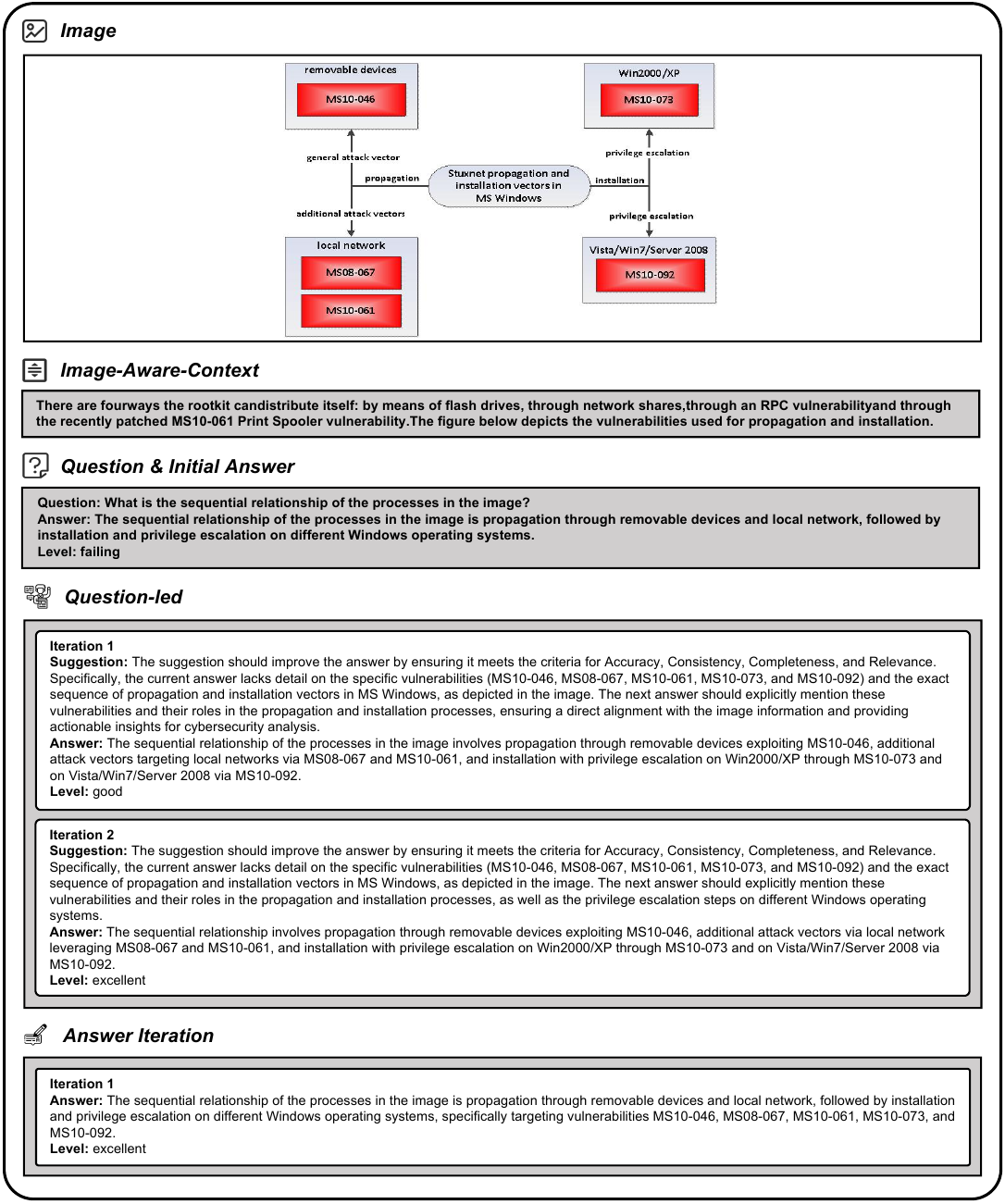}
    \caption{Iteration Example}
    \label{IterationExample}
\end{figure*}

\begin{figure*}[!t]
    \centering
    \includegraphics[width=0.95\linewidth]{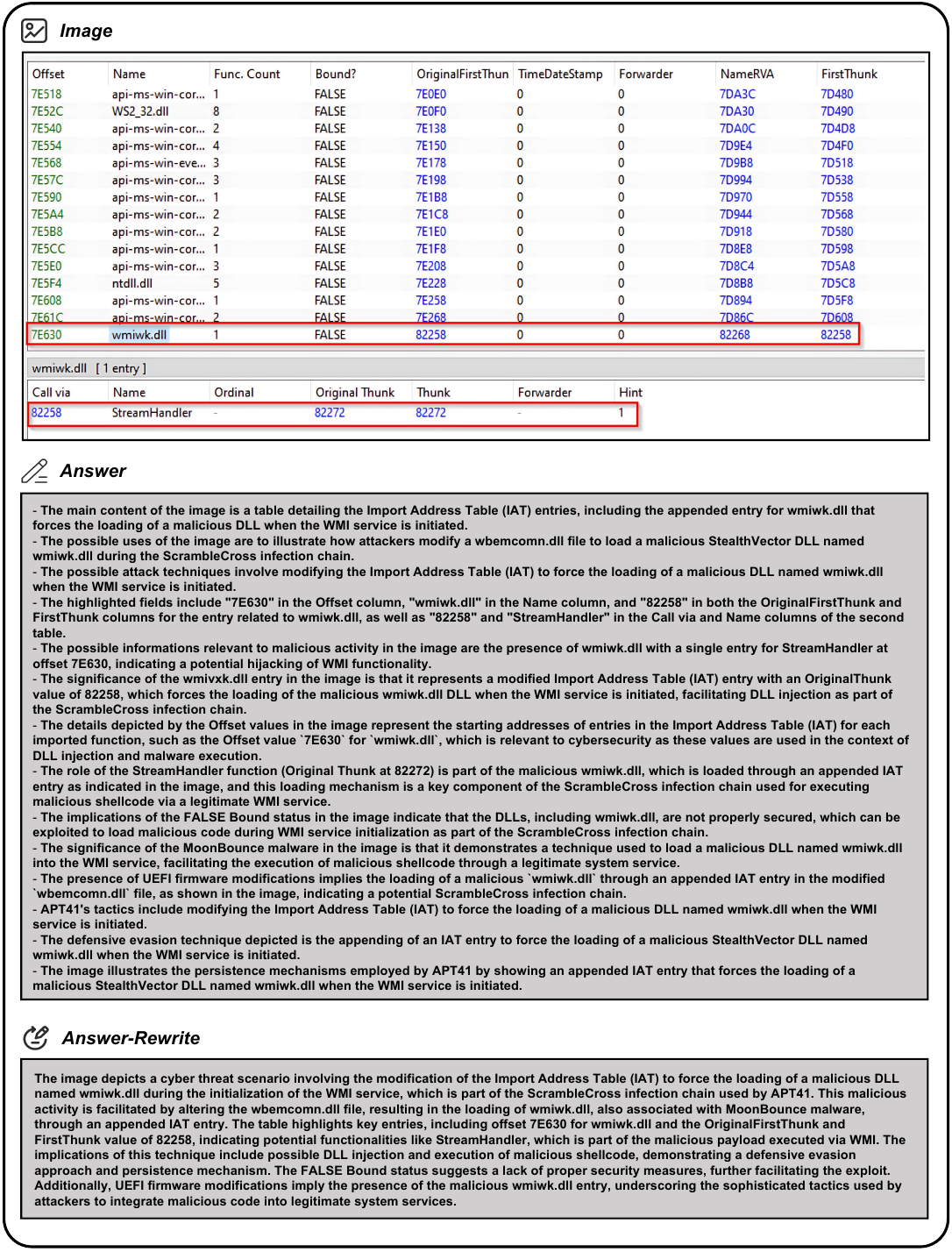}
    \caption{Combine Answer Example}
    \label{CombineAnswerExample}
\end{figure*}

Here we show sample tasks for each stage of the MM-AttacKG runtime: Figure~\ref{QA_Example} shows the question-answer process for the example image. Figure~\ref{AnswerExample} shows the differences in the answers for the example image with different textual support. Figure~\ref{IterationExample} shIterationExampleows the process of optimization of the answers. Figure~\ref{CombineAnswerExample} shows CombineAnswerExamplethe process of transforming the answers into reference combinatorial process

\section{}
\label{AB}
Here we show the LLMs prompt templates involved in the MM-AttacKG runtime: Table~\ref{tab:ques-seed} shows the related samples of leading Questions. Table~\ref{tab:image-prompt} shows the template of the task instruction prompt in different phases, which contain the generation of questions, the answer answering, the evaluation of the answers, the optimization of the answers.

\begin{table*}[h]
\centering
\caption{Leading Questions.}
\label{tab:ques-seed}
\begin{tabular}{m{3.5cm}|m{12cm}}
\hline
\multirow{5}{3.5cm}{\textbf{Attack Flow or Intelligence Structure}} 
& \rule{0pt}{2.5ex}- What is the main content of the image? \\
& - What are the possible uses of the image? \\
& - What are the possible attack techniques involved in the image? \\
& - What is the sequential relationship of the processes in the image? \\
& - What are the possible targets of attack in the image? \rule[-7pt]{0pt}{0pt}\\
\hline
\multirow{5}{3.5cm}{\textbf{Malware Code}} 
& \rule{0pt}{2.5ex}- What is the main content of code in the image? \\
& - What are the possible uses of the image? \\
& - What are the possible attack techniques involved in the image? \\
& - What is the possible function of the Code in the image? \\
& - What are the possible variables in the Code in the image? \rule[-7pt]{0pt}{0pt}\\
\hline
\multirow{4}{3.5cm}{\textbf{Application Tool Screenshot}} 
& \rule{0pt}{2.5ex}- What is the main content of the image? \\
& - What are the possible uses of the image? \\
& - What are the possible attack techniques involved in the image? \\
& - What is the key highlighted information in the picture? \rule[-7pt]{0pt}{0pt}\\
\hline
\multirow{5}{3.5cm}{\textbf{Data Table}} 
& \rule{0pt}{2.5ex}- What is the main content of the image? \\
& - What are the possible uses of the image? \\
& - What are the possible attack techniques involved in the image? \\
& - What are the fields highlighted in the image? \\
& - What are the possible informations relevant to malicious activity in the image? \rule[-7pt]{0pt}{0pt}\\
\hline
\multirow{4}{3.5cm}{\textbf{Charts and Data Visualization}} 
& \rule{0pt}{2.5ex}- What is the main content of the image? \\
& - What are the possible uses of the image? \\
& - What are the trends reflected in the image? \\
& - What are the conclusions that can be made based on the image? \rule[-7pt]{0pt}{0pt}\\
\hline
\multirow{4}{3.5cm}{\textbf{File Paths and Names}} 
& \rule{0pt}{2.5ex}- What is the main content of the image? \\
& - What are the possible uses of the image? \\
& - What are the possible attack techniques involved in the image? \\
& - What paths are included in the image? \rule[-7pt]{0pt}{0pt}\\
\hline
\multirow{3}{3.5cm}{\textbf{Descriptive Image and Content Explanation}} 
& \rule{0pt}{2.5ex}- What is the main content of the image? \\
& - What are the possible uses of the image? \\
& - What are the possible attack techniques involved in the image? \rule[-7pt]{0pt}{0pt}\\
\hline
\end{tabular}
\end{table*}

\begin{table*}[!t]
\centering
\caption{Prompts of image function identification module.}
\label{tab:image-prompt}
\scalebox{0.90}{
\begin{tabular}{m{3.5cm}|m{13.7cm}}
\hline
\multirow{11}{3.5cm}{\textbf{Question Generation}}& \rule{0pt}{2.5ex}You are a cybersecurity threat intelligence analyst. Please generate relevant questions derived from the content of an image based on the following rules: \\
& 1. Review the list of existing questions provided and generate new questions that explore different perspectives, details, or contexts within the image. \\
& 2. Generate new questions that explore different perspectives or details of the image based on the information from the knowledge graph. \\
& 3. These new questions should help further analyze the image from different perspectives. \\
& 4. These new questions should be related to cybersecurity or assist in the analysis of cyber threat intelligence. \\
& 5. Refer to the format of the given questions, which follows the pattern: "What is/are the XXX of/in the image?", "Where XXX should be replaced with a specific aspect of the image?". \rule[-7pt]{0pt}{0pt}\\
\hline
\multirow{8}{3.5cm}{\textbf{Question Answering}} 
& \rule{0pt}{2.5ex}You are a cybersecurity threat intelligence analyst. Please answer the questions based on the following rules: \\
& 1. Your answer must strictly adhere to the content visible in the image when mentioning any entities, objects, and their relationships. \\
& 2. Your answer must include a topic phrase that is specific to the question. \\
& 3. Your answer should be a single, concise sentence. \\
& 4. Only provide the direct answer to the question. Do not provide explanations or reasons for uncertainty. \rule[-7pt]{0pt}{0pt}\\
\hline
\multirow{20}{3.5cm}{\textbf{Answer Evaluation}} 
& \rule{0pt}{2.5ex}You are a cybersecurity threat intelligence analyst. Please rate the description based on the following rules: \\
& 1. Evaluate the description using the following four criteria:\\
    & - Accuracy: Accuracy represents whether the description accurately answers the question.\\
    & - Consistency: Consistency represents whether the description maintains content relevance to the image information.\\
    & - Completeness: Completeness represents whether the description adequately addresses the needs of the question.\\
    & - Relevance: Relevance represents whether the description is relevant to the cybersecurity field or useful for cyber threat analysis. \\
& 2. Apply the following rating scale based on the overall quality:\\
    & - "excellent": The description meets three of the criteria with only minor flaws or imperfections.\\
    & - "good": The description meets two of the criteria with small deviations or omissions that do not significantly impact the overall quality.\\
    & - "satisfactory": The description meets two of the criteria, but contains more noticeable flaws.\\
    & - "failing": The description meets only one of the criteria or none at all, with significant flaws that make the response unable to provide useful or relevant information. \\
& 3. If there are statements in the description such as unknown, no details, not mentioned, etc., mark it as "failing". \\
& 4. Your answer should be a single word: either "excellent", "good", "satisfactory", or "failing". \rule[-7pt]{0pt}{0pt}\\
\hline
\multirow{14}{3.5cm}{\textbf{Answer Optimization}} 
& \rule{0pt}{2.5ex}You are a cybersecurity threat intelligence analyst. Please provide your answers to the following questions again, using the image as a reference, based on the following rules: \\
& 1. Re-answer the questions to ensure the answers meet the following four criteria:\\
    & - Accuracy: Accuracy represents the answer accurately answers the question.\\
    & - Consistency: Consistency represents the answer maintains content relevance to the image information.\\
    & - Completeness: Completeness represents the answer adequately addresses the needs of the question.\\
    & - Relevance: Relevance represents the answer is relevant to the cybersecurity field or useful for cyber threat analysis.  \\
& 2. Improve existing unqualified answers (Paradigm 1) or re-answer questions based on suggestions provided (Paradigm 2). \\
& 3. Ensure the revised answer differs from the previous unqualified answer (Paradigm 1), or strictly follows the suggestions given (Paradigm 2). \rule[-7pt]{0pt}{0pt}\\
\hline
\end{tabular}
}
\end{table*}

\begin{table*}[!t]
\centering
\caption{Prompts of ICL.}
\label{tab:image-prompt-ICL}
\scalebox{0.90}{
\begin{tabular}{m{3.5cm}|m{13.7cm}}
\hline
\multirow{12}{3.5cm}{\textbf{Extraction of Entities and Relation}}& \rule{0pt}{2.5ex}You are a cyber-security information-extraction specialist. Given an image, its context, and a CTI summary, identify entities and their relationships, then output them as triplets:
Subject (Type); Relation; Object (Type). \\
& [Input] \\
& 1. An image; 2. Image In-context; 3. CTI summary; 4. Entity Types \& Descriptions; 5. Relation Table.\\
& [Example]\\
& Input: \\
& Image: ... attack path from public Internet → SSH→ Public-facing Server1...\\
& In-context: This led our responders to identify the occurrence of...\\
& CTI summary: CrowdStrike’s analysis of the StellarParticle campaign...\\
& Output:\\
& Public Internet(infrastructure); communicate-with; Public-facing Server1(infrastructure)\\
& ...\rule[-7pt]{0pt}{0pt}\\
\hline
\multirow{11}{3.5cm}{\textbf{Extraction of Techniques}} 
& \rule{0pt}{2.5ex}You are a cyber-threat intelligence specialist. Given an image, its context, and a CTI summary, identify the one MITRE ATT\&CK tactic that best matches the image.\\
& [Input]\\
& 1. CTI image; 2. Image context; 3. CTI summary; 4. MITRE ATT\&CK tactics list.\\
& [Example]\\ 
& Input: \\
& Image: ... attack path from public Internet → SSH→ Public-facing Server1...\\
& In-context: This led our responders to identify the occurrence of...\\
& CTI summary: CrowdStrike’s analysis of the StellarParticle campaign...\\
& Output:\\
& Lateral Movement \rule[-7pt]{0pt}{0pt}\\
\hline
\multirow{20}{3.5cm}{\textbf{Integrated Attack Graph}} 
& \rule{0pt}{2.5ex}You are a cyber-threat intelligence specialist. Given a CTI image, and a Knowledge Graph of existing triplets, perform one of three operations based on the image:\\
& 1. New Node Addition: add a new entity to extend an existing triplet.\\
& 2. New Relationship Addition: link entities from different triplets.\\
& 3. Technique Addition: tag an existing triplet with a new MITRE ATT\&CK technique.\\
& [Input]\\
& 1. CTI image; 2.Knowledge Graph (list of triplets); 3. Entity Types \& Descriptions; 4.MITRE ATT\&CK techniques list.\\
& [Rules]\\
& Rule 1: New Node Addition: if the image shows an entity related to a KG triplet, output a JSON list [...] with objects containing:\\
& ...\\
& description: reason for adding the node\\
& new\_node: \{id, type, properties:\{description\}\}\\
& relationship: \{Subject, SubjectType, Relation, Object, ObjectType\}\\
& Rule 2: New Relationship Addition: if the image connects entities from different triplets, output objects with:\\
& ...\\
& description: reason for the new relation\\
& relationship: \{Subject, SubjectType, Relation, Object, ObjectType\}\\
& Rule 3: Technique Addition: if the image indicates a new MITRE technique for a triplet, output objects with:\\
& ...\\
& [Example]\\
& Input:\\
& Image: attacker executes Stuxnet injection, installs a malicious module...\\
& Knowledge Graph: triplets for Stuxnet → install → dropper; attackers → develop → capabilities...\\
& Entity Types: includes campaign...\\
& MITRE list: includes T1055-Process Injection...\\
& Output:\\
& [\\
&   \{"type":"new\_node\_addition", ...\},\\
&   \{"type":"new\_relationship\_addition", ...\},\\
&   \{"type":"technique\_addition", ...\\\
& ]\rule[-7pt]{0pt}{0pt}\\
\hline
\end{tabular}
}
\end{table*}

\begin{table*}[!t]
\centering
\caption{Prompts of CoT.}
\label{tab:image-prompt-CoT}
\scalebox{0.90}{
\begin{tabular}{m{3.5cm}|m{13.7cm}}
\hline
\multirow{6}{3.5cm}{\textbf{Extraction of Entities and Relation}}& \rule{0pt}{2.5ex}You are a cybersecurity information‐extraction specialist. Given a CTI image (plus its context and summary for reference), identify visible entities and their relationships, then output them as triplets:
Subject(type); relation; Object(type).\\
& [Thinking Process]\\
& 1. Spot entities in the image.\\
& 2. Map each to the shortest matching provided type.\\
& 3. Identify clear, active‐voice relationships between them.\\
& 4. Form triplets.\\
\hline
\multirow{7}{3.5cm}{\textbf{Extraction of Techniques}}& \rule{0pt}{2.5ex}You are a cyber-threat intelligence expert. Given a CTI image (with optional context/summary) and a list of MITRE ATT\&CK tactics, identify the single tactic that best matches the image.\\
& [Thinking Process]\\
& 1. Inspect the image for attack indicators (flowcharts, logs, interfaces).\\
& 2. Identify candidate tactics from the provided list.\\
& 3. Select the one tactic that most directly reflects what you see.\\
& 4. Ensure it fits the attack phase implied by the image.\\
\hline
\multirow{35}{3.5cm}{\textbf{Integrated Attack Graph}} 
& \rule{0pt}{2.5ex}You are a cyber-threat intelligence expert. Given a CTI image, a Knowledge Graph of triplets, Entity Types \& Descriptions, and MITRE ATT\&CK techniques, extend the graph by discovering:\\
& 1. New Node Addition: adding an image-derived entity to an existing triplet.\\
& 2. New Relationship Addition: linking entities from different triplets.\\
& 3. Technique Addition: tagging a triplet with a new MITRE technique.\\
& [Thinking Process]\\
& 1. Identify triplets strongly correlated with the image.\\
& 2. Extract and type entities visible in the image.\\
& 3. For each strong-match triplet: check if an image entity adds a new connection to its subject or object (new node addition).\\
& 4. Check if an entity in this triplet should link to an entity in another triplet (new relationship addition).\\
& 5. Determine if the image implies a new MITRE technique for the most relevant triplet (technique addition).\\
& 6. Assemble JSON outputs.\\
& 7. If none apply, output No Match.\rule[-7pt]{0pt}{0pt}\\
& [Rules]\\
& Rule 1: New Node Addition: if the image shows an entity related to a KG triplet, output a JSON list [...] with objects containing:\\
& ...\\
& description: reason for adding the node\\
& new\_node: \{id, type, properties:\{description\}\}\\
& relationship: \{Subject, SubjectType, Relation, Object, ObjectType\}\\
& Rule 2: New Relationship Addition: if the image connects entities from different triplets, output objects with:\\
& ...\\
& description: reason for the new relation\\
& relationship: \{Subject, SubjectType, Relation, Object, ObjectType\}\\
& Rule 3: Technique Addition: if the image indicates a new MITRE technique for a triplet, output objects with:\\
& ...\\
& description: reason for the new technique\\
& target\_relationship: \{Subject, Relation, Object\}\\
& new\_techniques: ["technique\_id - technique\_name"]\\
& Rule 4: Use active‑voice, concise verb phrases for relations.\\
& Rule 5: Only generate JSON if there is a strong match; otherwise output No Match.\rule[-7pt]{0pt}{0pt}\\
\hline
\end{tabular}
}
\end{table*}

\end{document}